\documentclass[5p]{elsarticle}

\usepackage{booktabs} % For formal tables
\usepackage{listings}
\usepackage{changepage}
\usepackage{url}
\usepackage{interval}

\usepackage{amssymb} %for \checkmark

\usepackage[ruled,linesnumbered]{algorithm2e} % For algorithms

\SetAlFnt{\small}
\SetAlCapFnt{\small}
\SetAlCapNameFnt{\small}
\SetAlCapHSkip{0pt}
\IncMargin{-\parindent}

\def\eg{e.\,g.}
\def\ie{i.\,e.}
\begin{document}
\begin{frontmatter}

%\title{Dynamic Autotuning of CUDA and OpenCL Kernels with Kernel Tuning Toolkit}
\title{A Benchmark Set of Highly-efficient CUDA and OpenCL Kernels and its Dynamic Autotuning with Kernel Tuning Toolkit}

\author[ics]{Filip Petrovi\v{c}}
\ead{fillo@mail.muni.cz}
\author[ics,cnb]{David St\v{r}el\'{a}k}
\ead{373911@mail.muni.cz}
\author[ics]{Jana Hozzov\'{a}}
\ead{hozzova@mail.muni.cz}
\author[ics]{Jaroslav Ol'ha}
\ead{348646@mail.muni.cz}
\author[ics]{Richard Trembeck\'{y}}
\ead{422536@mail.muni.cz}
\author[vie]{Siegfried Benkner}
\ead{siegfried.benkner@univie.ac.at}
\author[ics,vie]{Ji\v{r}\'{i} Filipovi\v{c}\corref{mycorrespondingauthor}}
\ead{fila@mail.muni.cz}
\cortext[mycorrespondingauthor]{Corresponding author\\ \copyright 2020. This manuscript version is made available under the CC-BY-NC-ND 4.0 license http://creativecommons.org/licenses/by-nc-nd/4.0}
\address[ics]{Institute of Computer Science, Masaryk University, Botanick\'{a} 68a, 60200 Brno, Czech Republic}
\address[cnb]{Spanish National Centre for Biotechnology, Spanish National Research Council, Calle Darwin, 3, 28049 Madrid, Spain}
\address[vie]{Faculty of Computer Science, University of Vienna, W\"{a}hringer Str. 29, Vienna 1090, Austria}

\begin{abstract}
In recent years, the heterogeneity of both commodity and supercomputers hardware has increased sharply. Accelerators, such as GPUs or Intel Xeon Phi co-processors, are often key to improving speed and energy efficiency of highly-parallel codes. However, due to the complexity of heterogeneous architectures, optimization of codes for a certain type of architecture as well as porting codes across different architectures, while maintaining a comparable level of performance, can be extremely challenging.
%Even if a code has been written in such a way that it can be ported across different architectures, such functional portability usually does not imply performance portability.
% On the contrary, the performance of highly optimized code is rarely portable between hardware architectures, or even the same hardware when computation settings or input data change, without significant re-tuning.
%Autotuning is a method for automatically exploring combinations of tuning parameters: properties of the application and system which are supposed to influence performance. 
Addressing the challenges associated with performance optimization and performance portability, autotuning has gained a lot of interest. Autotuning of performance-relevant source-code parameters allows to automatically tune applications without hard coding optimizations and thus helps with keeping the performance portable. %Autotuning can be performed offline during design or installation time, or dynamically during the application runtime.
%SB usually there are no seperate paragraphs (line-breask) in an abstract
In this paper, we introduce a benchmark set of ten autotunable kernels 
for important computational problems
implemented
in OpenCL or CUDA. Using our Kernel Tuning Toolkit, we show that with autotuning most of the kernels reach near-peak performance on various GPUs and outperform baseline implementations on CPUs and Xeon Phis. Our  evaluation also demonstrates that autotuning is key to performance portability.
%The performance portability of the kernels is provided by code optimization parameters autotuning. 
%
In addition to offline tuning, we also introduce dynamic autotuning of code optimization parameters during application runtime. With dynamic tuning, the Kernel Tuning Toolkit enables applications to re-tune performance-critical kernels at runtime whenever needed, for example, when input data changes. Although it is generally believed that autotuning spaces % are extremely large and thus difficult to search 
tend to be too large to be searched during application runtime, we show that it is not necessarily the case when tuning spaces are designed rationally. % (\ie{}, by kernel developer). 
Many of our kernels reach near peak-performance with moderately sized tuning spaces that can be searched at runtime with acceptable overhead. 
%Therefore, dynamic autotuning %has acceptable overhead, and it can improve the speed of programs with even a short runtime. 
Finally we demonstrate, how dynamic performance tuning can be integrated into a real-world application from cryo-electron microscopy domain.

%On a set of ten benchmarks, we demonstrate that autotuning OpenCL and CUDA code optimizations can be used to reach near-peak performance on a variety on GPU hardware and significantly improve the performance of a code ported on CPU and MIC. Moreover, we show that in many cases the space of the tuning parameters is not too hard to search and therefore dynamic autotuning can improve the speed of programs with even a short runtime. The integration of the Kernel Tuning Toolkit performing dynamic tuning is demonstrated on example of a real-world parallel application from cryo-electron microscopy domain.

%We have applied autotuning to 10 kernels relevant in high performance computing applications and we have evaluated them on multiple hardware devices of different architectures. Although it is generally believed that autotuning spaces are extremely large and thus difficult to search during application runtime, we show that is not necessary the case when tuning spaces are designed rationally (\ie{} by kernel developer). We demonstrate that our kernels can be tuned to high performance close to hardware theoretical peak, or at least outperforming the baseline implementation, whereas tuning spaces are of reasonable size allowing dynamic tuning in many cases. Last but not least, we demonstrate that Kernel Tuning Toolkit can be integrated into complex parallel application and perform dynamic autotuning transparently.
\end{abstract}

%TODO highliths:
% - We introduce Kernel Tuning Toolkit, an easy-to-use tool allowing full integration into application code and transparent dynamic tuning of OpenCL or CUDA kernels.
% - We demonstrate that rationally designed tuning spaces can be searched during runtime with reasonable overhead yielding significant improvements of performance portability.
% - We show integration of Kernel Tuning Toolkit into 3D Fourier reconstruction application in Xmipp suite.


\end{frontmatter}	

\section{Introduction}
In recent years, the acceleration of complex computations using hardware accelerators have become much more common. Currently, there are many devices developed by multiple vendors which differ in hardware architecture, performance, and other attributes. In order to support application development for these devices, several APIs such as OpenCL (Open Computing Language) or CUDA (Compute Unified Device Architecture) were designed. A code written in those APIs is functionally portable: it can be executed on various devices while producing the same result. However, performance portability is often  limited due to the different hardware characteristics of these devices. For example, an OpenCL code which was optimized for a GPU may perform poorly on a CPU and vice versa. The performance portability issues may even exist among different generations of devices developed by the same vendor~\cite{kurzak2012autotuning}. Moreover, code performance may be sensitive to input size, structure, or application settings, so a code optimized for some input may run sub-optimally when the input is changed~\cite{gonzalo2017revisiting, strelak2019gpu}.

A costly solution to this problem is to manually optimize code for each utilized device and possibly also for multiple sizes or structures of the input. An alternative solution is a technique called \textit{autotuning}. Autotuning allows optimizing the application's \textit{tuning parameters} (properties influencing the application performance) in order to perform the execution more efficiently. It is a general technique with a broad range of applications, which includes areas such as network protocols, compilers, and database systems. We focus on \textit{autotuning of code optimization parameters}, which allows changing the application at the level of its source code: %how the application is implemented: 
from low-level optimizations such as loop-tiling or unrolling to more aggressive changes such as modification of data layout or even using a different algorithm.

In this paper, we introduce the Kernel Tuning Toolkit (KTT), which focuses on autotuning of codes written in OpenCL or CUDA. With KTT, tuning parameters change the source code in a way defined by a programmer via preprocessor macros. Thus, tuning parameters may affect virtually any property of the source code, making autotuning very powerful. KTT targets expert programmers, as potential code optimizations have to be implemented explicitly, requiring detailed knowledge of hardware architectures. %It is a universal but low-level alternative to a code variant tuning (entire pre-implemented functions are altered) and compiler-based autotuning (the automatic transformations of the code are employed). Therefore, one class of KTT users are expert programmers, who are able to identify tuning parameters and implement their effect in the OpenCL or CUDA source code. Another class is application programmers using a library of autotuned kernels, who may need to call KTT API, but are not required to have a background in a hardware-aware optimization for accelerators.

%Although benefits of autotuning have been proved many times, it is still far from becoming mainstream in high performance computing. The only exception are autotuned libraries, such as ATLAS~\cite{whaley1998automatically} or FFTW~\cite{frigo2005design}, when autotuning is used just for the library functions. 

Autotuning can be performed \textit{offline}\footnote{We adopt the nomenclature from~\cite{balaprakash2018autotuning}.}, \ie{}, before the execution of a tuned code. Offline tuning is easier to implement but does not allow an application to re-tune when its environment changes. \textit{Online} autotuning allows the application to tune itself during runtime by means of changing some runtime parameters. With \textit{dynamic} autotuning, the application can even build the space of different variants during runtime, \ie{}, it is able to compile tuned kernels during the tuning process. Although several code parameters  autotuning frameworks for heterogeneous computing have been introduced~\cite{ansel2014opentuner, nugteren2015cltune, rasch2018atf, vanwerkhoven2018kernel}, they are intended to be used in a \textit{standalone tuning tool}, supporting offline autotuning only. On the other hand, KTT can be integrated into application code and supports also dynamic tuning.

A tighter integration into applications has been recently identified as one of the main challenges in autotuning~\cite{balaprakash2018autotuning}. Kernel Tuning Toolkit was designed to simplify the integration process. It acts as an intermediate layer between the application and OpenCL or CUDA API. Therefore, the application source code has to be adapted to incorporate KTT calls. However, once integrated, the application can transparently switch between execution and tuning of the kernels. For example, the application can re-tune itself if it is executed on new hardware, or start its execution with already optimized tuning parameters, and automatically start re-tuning during runtime when the input changes.

Using KTT, we have developed a benchmark set comprising ten autotuned codes. We have executed the benchmark set on multiple hardware devices, including GPUs from NVIDIA and AMD, CPU and the Xeon Phi. We prove that our autotuned implementations are efficient enough -- they often reach performance close to the theoretical peak of the hardware or at least outperform the baseline (\ie{}, not autotuned) implementations significantly. We also show that autotuning is required to ensure performance portability of the codes. 

The search for efficient tuning configurations may be challenging due to the discrete and non-linear nature of tuning spaces~\cite{balaprakash2018autotuning}. Therefore, large tuning spaces are usually impossible to explore during application runtime. 
%However, we argue here that with code optimization autotuning, the tuning spaces are created rationally (\ie{}, by a programmer) and thus may have rather a moderate size. 
However, if tuning spaces a are created rationally (\ie{}, by an expert programmer), their exploration may be feasible even at runtime. 
Expert programmers have to understand the effect of tuning parameters %to be able to implement them 
and %thus can 
set reasonable boundaries to their values. For example, setting the acceptable sizes of work-groups to multiples of 32, as it is suitable for vectorization on CPUs and Xeon Phis and efficient on GPUs executing work-items in warps. We show in this paper that rationally constructed tuning spaces can be moderately-sized  (thousands of configurations or less) and still contain enough good configurations required for performance portability. Such tuning spaces can be searched during application runtime without too high overhead. To prove the applicability of KTT in real applications, we demonstrate dynamic autotuning in a CUDA-accelerated 3D Fourier Reconstruction in Xmipp~\cite{strelak2019gpu}.% The 3D Fourier Reconstruction uses multiple CPU threads, hybrid CPU-GPU computation and asynchronous kernel execution and memory copy, which needs to be reflected in code with integrated autotuner.

The paper makes the following major contributions:
\begin{itemize}
  \item \textit{Development of dynamic autotuning techniques in the Kernel Tuning Toolkit.} KTT introduces a high-level API for kernels and data manipulation, which can be easily used and integrated into applications. It allows % to % https://english.stackexchange.com/a/196130
 switching transparently between autotuning and executing tuned kernels. KTT is open-source\footnote{https://github.com/Fillo7/KTT}, fully documented, and contains many examples of its usage.
  \item \textit{Introduction of a benchmark set of autotuned kernels.} We have conducted a benchmark set, including multiple kernels relevant for HPC, 
spanning across multiple application domains such as image processing, linear algebra, computational chemistry, and differential equations.
  We demonstrate that autotuning of optimization parameters  improves the performance portability of the benchmark set across a range of different heterogeneous architectures significantly. We also show that rationally constructed tuning spaces can be searched fast enough to allow dynamic tuning in many cases.
  \item \textit{Demonstration of dynamic tuning with a real-word application.} We show that dynamic autotuning can be used in a real-world application, such as a 3D Fourier Reconstruction. Dynamic autotuning is also demonstrated on an application performing batched matrix multiplication with varying matrix sizes. We experimentally evaluate the speed of tuning space search convergence as well as dynamic tuning overhead on these examples.
\end{itemize}

The rest of the paper is organized as follows. In Section~\ref{sect:related_work}, we introduce related work and compare it with our work. The main design decisions and concepts of KTT are described in Section~\ref{sect:design}. Section~\ref{sect:benchmarks} introduces a set of ten autotunable benchmarks and evaluates their efficiency and performance portability. Dynamic autotuning is evaluated in Section~\ref{sect:dynamic_eval}. We conclude and sketch future work in Section~\ref{sect:conclusion}.

\section{Related Work}
\label{sect:related_work}

In this section, we compare our work to state-of-the art methods in autotuning in three areas: tuning targets (which properties are tuned), tuning time (when tuning is performed) and search strategies (how the tuning space is searched and evaluated).

Autotuning covers a broad range of empirically tuned parameters related to application performance, such as compiler parameters, or the runtime environment~\cite{gerndt2015automatic, gerndt2017multi}. Some autotuners do not required to modify the application source code, for example, compiler flags tuners~\cite{bajrovic2016tuning} or MPI tuner~\cite{miceli2013autotune}. Other tuners may change application source code in order to test different code optimization variants. We focus on the autotuners altering the code of applications in the rest of this section as they are directly related to our tuner.

Autotuning is already successfully deployed in some high-performance libraries for conventional CPUs, such as ATLAS~\cite{whaley1998automatically} (linear algebra) or FFTW~\cite{frigo2005design} (signal processing). %cuFFTAdvisor? 
Libraries for accelerators are also often improved by autotuning~\cite{li2009note, grewe2011automatically, li2013mpfft, matsumoto2012performance}. However, those libraries use autotuners specially designed for them. Here, we are interested in generic autotuners. %tuners are designed especially for those libraries. 
Frameworks for skeletons or DSLs also use autotuning to search for the best combination of the implementation variants empirically~\cite{enmyren2010towards, gray2012autotuning, filipovic2013optimizing, bajrovic2014automatic}. While they cover a broader range of applications compared to autotuned libraries, they are still restricted to a particular problem domain or a set of skeletons. 

Code optimizations autotuners generate multiple func\-ti\-on\-ally-equivalent variants of the application source code.
% variants from user-instrumented source code~\cite{nelson2008model, tiwary2011online, nugteren2015cltune, falch2015machine} \todo{zkontrolovat ty clanky, jestli tam neni neco compiler-based}. 
They may select one of the predefined variants of a tuned function~\cite{muralidharan2014nitro}, or generate and compile implementations according to the values of the tuning parameters. We distinguish between compiler-based tuning, where the space of code transformation is generated automatically~\cite{rudy2011programming, tiwary2011online, steuwer2017lift} and user-defined code optimization parameters autotuning~\cite{nugteren2015cltune, falch2015machine, rasch2018atf, filipovic2017autotuning}. User-defined code optimization parameters tuning requires expert programmers to identify and implement tuning possibilities in the source code manually (\eg{}, by using preprocessor macros). Even though this approach may be costly in terms of time and expertise of the programmer, it allows to explore highly diversified variants of the code, which usually cannot be generated automatically by compilers: the programmer can change virtually anything, for example, alter algorithms (\eg{}, use merge sort instead of quicksort) or change the data layout in the memory (\eg{}, use a structure of arrays instead of an array of structures). %Moreover, as the expert programmer knows hardware a design of the tuning parameters is up to the programmer, tuning spaces are usually smaller comparing to those generated by compilers.

Our Kernel Tuning Toolkit focuses on tuning of user-defined code optimization parameters. Most similar to our work are CLTune~\cite{nugteren2015cltune}, AUMA~\cite{falch2015machine}, ATF~\cite{rasch2017atf, rasch2018atf}, and Kernel Tuner~\cite{vanwerkhoven2018kernel}, which are problem domain-agnostic autotuners designed for heterogeneous computing. %All those tuners are able to propagate tuning parameters as preprocessor macros, but excepting Kernel Tuner, the parameters may affect one kernel only. 

Existing benchmark sets for heterogeneous computing, such as Parboil~\cite{stratton2012parboil}, SHOC~\cite{danalis2010scalable}, or Polybench/GPU~\cite{gray2012polybench} do not support autotuning of code optimization parameters (only work-group size can be typically changed without substantial rewriting the benchmark). To the best of our knowledge, there is no comprehensive benchmark set for code optimization parameters tuning in heterogeneous computing. In~\cite{nugteren2015cltune, rasch2018atf}, two benchmarks are used to evaluate code optimization parameters tuning: GEMM and 2D convolutions. Those benchmarks are also used in our benchmark set. Three benchmarks are used in~\cite{vanwerkhoven2018kernel} (one of them is the GEMM introduced in~\cite{nugteren2015cltune}) and in~\cite{falch2015machine}. In our work, a set of ten benchmarks is introduced.

Some forms of dynamic autotuning are supported by problem-specific autotuners, such as SpMV tuning~\cite{gonzalo2017revisiting} or generic autotuners, such as Active Harmony~\cite{tiwary2011online}. Autotuners may also support online autotuning where usually multiple variants of code are produced in an offline phase and searched during runtime. Online tuning is easier to implement than dynamic tuning (there is no runtime compilation), but it is not practical when the number of possible code variants is high. An examples of an  online tuner is SOCRATES~\cite{gadioli2018socrates, muralidharan2016architecture}. 

None of the frameworks for code optimizations in heterogeneous computing support dynamic tuning natively~\cite{nugteren2015cltune, falch2015machine, rasch2018atf, vanwerkhoven2018kernel}. To implement dynamic tuning with those frameworks, the programmer has to add a non-trivial amount of glue code, running the tuner during application runtime to find a better tuning configuration and then exporting this configuration into the application, typically by re-compiling OpenCL or CUDA kernels with the JIT compiler. The OpenTuner~\cite{ansel2014opentuner}, another similar tuner, is a more generic and low-level tool: it allows us to tune virtually any property of the application, but a higher amount of user effort has to be invested into the integration of the tuner. OpenTuner could be used for dynamic autotuning with higher effort than~\cite{nugteren2015cltune, falch2015machine, rasch2018atf, vanwerkhoven2018kernel}, since a code responsible for a tuned kernel compilation, execution, and testing has to be provided as well. On the other hand, OpenTuner would allow to use results computed by kernels during tuning, which can increase the performance of the tuned application. To the best of our knowledge, KTT is the first autotuning framework combining universal code optimization parameters tuning with native support of dynamic autotuning for heterogeneous computing.

In this paper, we extend state-of-the-art general-purpose code optimizations autotuners for heterogeneous computing with dynamic tuning. Although the concept of dynamic autotuning is well-known, it requires an architecture that hides the OpenCL or CUDA API in order to switch implementation of kernels. In addition we contribute to the state-of-the-art in autotuning by introducing a benchmark set of autotunable codes, evaluating the efficiency and performance portability of the benchmarks, and assessing how difficult it is to amortize the overheads of dynamic tuning. %search spaces.

The space of tuning parameters can be very difficult to search: it is discrete, non-linear, and non-convex. Although a promising method has been recently published~\cite{vanwerkhoven2018kernel}, the majority of papers report that random search is often as efficient or even more efficient than more sophisticated search methods~\cite{kisuki2000combined, seymour2008comparison, nugteren2015cltune}. Therefore, it can be difficult to search large tuning spaces containing hundreds of thousands of configurations or more. Extremely large tuning spaces, however, result mainly from compiler-based autotuners such as Lift~\cite{steuwer2017lift} or naively constructed tuning spaces. The papers~\cite{kisuki2000combined, seymour2008comparison, nugteren2015cltune, vanwerkhoven2018kernel} focus on the analysis of tuning space search methods. However, there has not been much effort invested into studying the size of tuning spaces using a larger number of benchmarks which maintain good performance portability across a wide range of different hardware architectures. In this paper, we constructed a set of ten benchmarks and show that tuning spaces are often small enough to be searched dynamically while still providing performance portability with a near-peak performance. 

Machine learning on historical autotuning data can be used to decrease the number of tuning decisions performed during program compilation or execution. In~\cite{muralidharan2016architecture}, a dynamic selection from a very limited number of code variants is based on a model created from previous tuning runs. In~\cite{cummins2017end}, a single tuning parameter can be optimized at compilation time by a neural network %learned from
trained in multiple trial runs. Contrary to those papers, we focus on multi-dimensional tuning spaces.

\section{Architecture of the Kernel Tuning Toolkit}
\label{sect:design}
In this section, we introduce the main architectural concepts and the API of the Kernel Tuning Toolkit. We are using the following terminology in the paper. A \textit{tuning parameter} is a variable which affects the code in a user-defined way (\eg{}, determines loop unroll factor). The \textit{tuning space} is a cross product of all the possible values of all tuning parameters. A \textit{configuration} is a single point in the tuning space (\ie{}, assignment of concrete values to all tuning parameters), which fully determines one possible implementation variant of the tuned code. The main functionality of KTT is:
\begin{itemize}
  \item specification of tuning parameters and constraints of tuning space;
  \item compiling and executing the kernel or \textit{kernel composition} (multiple kernels and host code with shared tuning parameters);
  \item automatically searching the tuning space;
  \item managing data transfers automatically (KTT automatically creates and copies data from/to the accelerator);
  \item checking the results of the tuned kernel against a reference implementation computation.
\end{itemize}

\begin{figure}[t]
    \centering
    \includegraphics[width=.75\hsize]{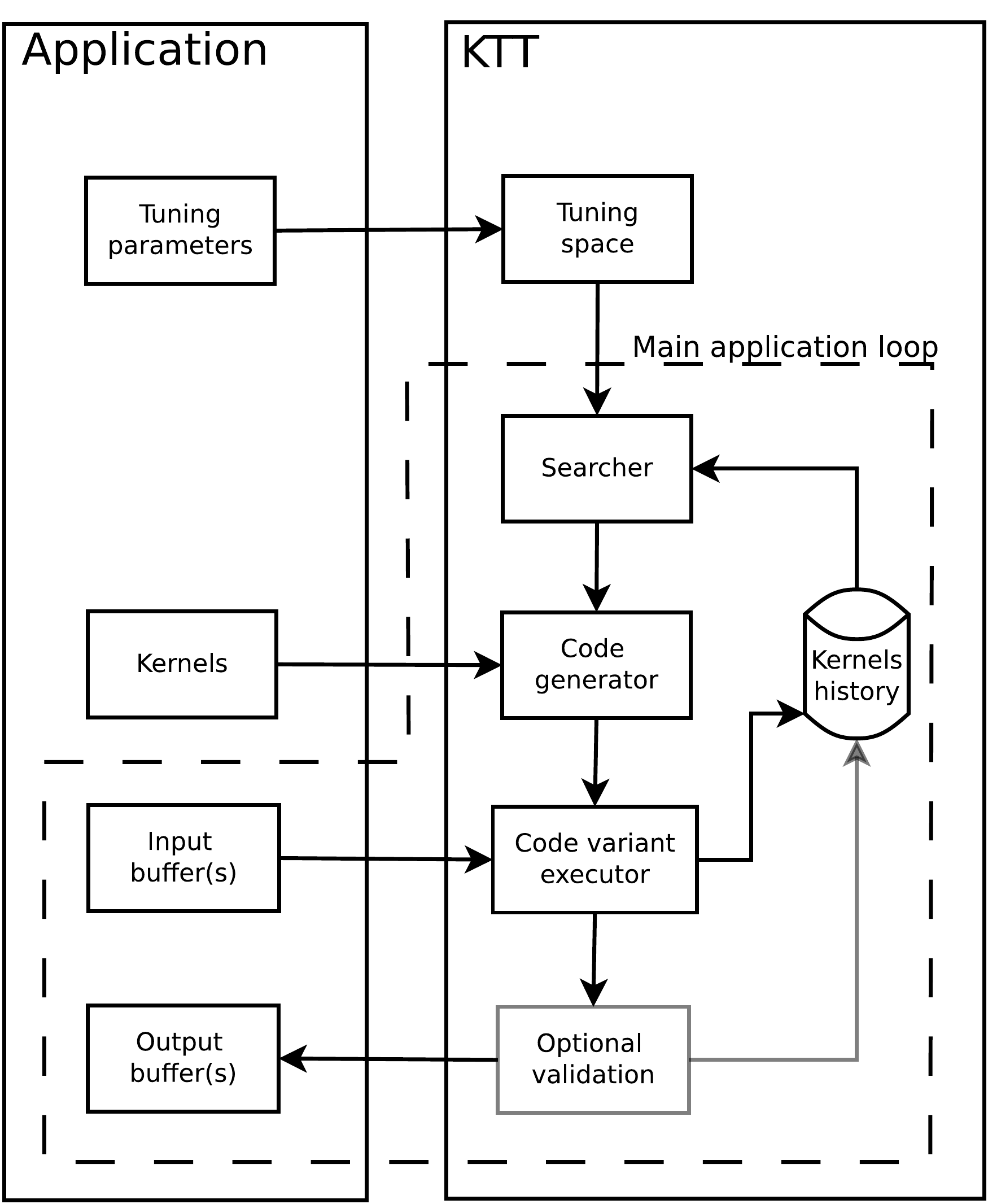}
    \caption{Schematic view of KTT architecture. The dashed line shows components, which are typically active during dynamic tuning inside the main application loop.}
    \label{fig:ktt-architecture}
\end{figure}

KTT has been designed as a C++ library, which replaces direct access to the OpenCL or CUDA API. By providing a middle layer between the application and the OpenCL or CUDA API, KTT is able to perform autotuning transparently: the kernel execution and tuning can be performed by the same application code. %it can iteratively recompile kernels, reload their inputs, or automatically check the correctness of outputs. 
However, in order to allow integration into real-world applications, KTT must support important functionality such as memory management, kernel configuration, execution, and synchronization provided by OpenCL or CUDA. 
Because KTT forms a middle layer between the application and the CUDA or OpenCL API, it can modify kernel code at runtime, transparently to the application. Moreover, this design allows switching between OpenCL and CUDA easily. When kernel codes for both APIs are provided by the programmer, the KTT is just initialized with the selected API and handles all the communication between the application and OpenCL or CUDA. Because OpenCL and CUDA use a different way to configure the parallelism of the kernel\footnote{OpenCL's NDRange describes global parallelism, whereas CUDA blocks and threads define different layers of parallelism}, KTT can automatically translate parallelism configuration for the selected API. The KTT API has been derived from the CLTune project~\cite{nugteren2015cltune}, so it is very similar to CLTune when we use it for offline tuning. Additionally, KTT API allows for tuning compositions of multiple kernels, tuning of how  kernels are called from host code~\cite{filipovic2017autotuning}, and novel features for dynamic tuning. %SB? what does ``or a host code'' mean? #: is it more clear now?

The architecture of KTT and its connection to the autotuned application is sketched in Figure~\ref{fig:ktt-architecture}. The application creates kernels and defines tuning parameters and their acceptable values (with possible constraints passed as lambda functions), and passes them to KTT, where the tuning space is built. Then, it connects input and output buffers to the kernels and starts the tuning process. 
%During a kernel(s) tuning, 
KTT uses a searcher to search the tuning space and to select a configuration to be executed. In current implementation, random search, simulated annealing and Markov-chain Monte Carlo searchers are available.
%SB? The supported search strategies could be briefly mentioned here
Then, it compiles the kernel(s) according to the selected configuration, executes and benchmarks it. If dynamic tuning is active, the results of the tuned kernel(s) can be immediately used by the application. The results can be validated against a reference implementation by KTT. The execution of kernel(s) is benchmarked and the performance results are stored in KTT, allowing the searcher to navigate the search process and the application to query for, \eg{}, the fastest configuration.
%SB? need to briefly describe here what the figure shows %JF better now? %SB yes! but available search strategies could be mentions

\subsection{Kernel tuning}
The simplest scenario %which is also supported by all of the tuners compared in Table~\ref{tab:comparison} 
is tuning of a single kernel. In this case, the following steps have to be done in a tunable code:
\begin{itemize}
  \item initialize the tuner;
  \item create handlers for kernel arguments;
  \item create the kernel handler;
  \item assign input/output arguments to the kernel;
  \item define tuning parameters, their acceptable values, and constraints;
  \item start tuning.
\end{itemize}
The tuner executes and benchmarks different tuning configurations and searches for the one, which results in the shortest kernel runtime. %It is up to the user to transfer those values into application code.

In many real-world applications, some tuning parameters are shared between multiple kernels (\eg{}, the memory layout of some intermediate data). KTT framework allows sharing tuning parameters among kernels by using \emph{kernel compositions}. Moreover, a portion of a computation can be performed on the host %a host site 
(\eg{}, a tuning parameter may determine how many times a kernel is executed or if a host code performs some pre-computation). KTT  uses the \emph{tuning manipulator} when tuning parameters influence the host code. The tuning manipulator class enables users to customize a portion of the framework's code that is responsible for kernel execution and buffer management, and optionally can perform some part of the computation directly in the C++ host code. The tuning manipulator must implement a method \emph{launchComputation}, which can execute multiple kernels, perform computations in C++, and transfer data between host and device. Tuning manipulators and kernel compositions allow to use tuning parameters, which cannot be implemented when kernels are tuned separately: for example, it is possible to change a format of the intermediate data exchanged between multiple kernels. %Kernel composition is a set of kernels which are tuned together because they share some tuning parameters (\eg{} memory layout of some intermediate data). It has to have a tuning manipulator, which defines how are kernels executed.
%The API also provides methods for changing and updating kernel arguments as these can be different for individual kernel executions, which is in contrast to single kernel tuning where all the executions happen under the same conditions.

\subsection{Offline and Dynamic tuning}
KTT supports different types of autotuning depending on the time when tuning is performed and on the level of integration:
\begin{itemize}
  \item \textit{Offline autotuning} is performed prior to the execution of an application, usually by an extra utility. %The advantage of the offline tuning is an easier design of the autotuner -- its API can be simpler as there is no need to integrate tuner into the application. 
Offline tuning does not require integration of the autotuner into the application. The tuning utility can search for tuning parameters of the computationally most demanding application kernels and then exports values of those parameters to the build system. The disadvantage is that the tuning process cannot be easily repeated \textit{inside} the application, \ie{}, during application runtime.
  \item \textit{Dynamic autotuning} is performed during application runtime. When tuning parameters change application source code, it must be modified according to the actual values of the tuning parameters  and recompiled. The application can execute autotuning at any time, \eg{}, when it is executed on a new hardware device or when a performance-relevant characteristic of the processed data changes\footnote{With the current version of KTT, the decision about when to tune is always made by the application.}. 
  %SB? maybe the term ``blocking'' is somehow misleading - or should be better explained? %JF I don't insist on blocking/nonblocking term, but I simply don't have better idea -- is this additional explanation OK, or do you have any idea about better terminology? - %SB i think it should be clear now
Dynamic tuning can be performed in a blocking manner (the tuner tests several tuning configurations %combinations of tuning parameters 
and selects the best one; the results of kernel executions are not passed to the application during tuning) or non-blocking manner (the result of each tested kernel variant is immediately used by the application). With blocking autotuning, KTT automatically replicates input and output arrays, so there is no side effect caused by kernel results on the application. Non-blocking tuning is more suitable for interactive applications or complex parallel workloads with many dependent tasks, where a slow response of some component may be critical to the overall performance.
%  \item \textit{Blocking online autotuning} is tuning which is integrated into an application, but the tuning process is decoupled from a computation of an actual result. The application can execute autotuning at any time (\eg{} when is executed on a new hardware device or when a performance-related characteristic of the processed data changes), however, tuning process does not produce computation results. The blocking online autotuning requires more advanced support from a tuning framework, as values of the tuning parameters have to be integrated into the application during its runtime. Therefore, just-in-time (JIT) compilation may be required in case of the implementation parameters tuning.
%  \item \textit{Non-blocking online autotuning} is able to deliver computation results during the tuning process. Autotuning framework transparently changes application code according to the selected values of the tuning parameters, but the tuned code still delivers result for the application. The tuning has still some overhead (\eg{} JIT compiler invocations in case of the implementation parameters tuning), however, the overhead is lower than in case of the blocking online autotuning. The non-blocking autotuning is more suitable for an interactive applications or complex parallel workloads with many depending tasks, where a slow response of some component may be critical to overall performance. The tuning framework has to maintain the connection of the autotuned code with the rest of the application in this case.
\end{itemize}

% umi KTT tunit multi-GPU kod, navic treba pusteny ruznych gpu?

The Kernel Tuning Toolkit %implements all the methods mentioned above. It 
can be integrated into application code so that the application code manages memory objects and executes kernels via the KTT API instead of directly using OpenCL or CUDA. In such a case, the application decides if KTT changes values of the tuning parameters and recompiles kernels (tuning mode) or if KTT just executes the kernels (running mode). 
Furthermore, the kernels' result can be used by the application even during the tuning process (a non-blocking tuning described above), which improves application performance, especially when the tuning overhead is low (\ie{} kernels runtime dominates compilation runtime). %Note that with non-blocking tuning, application has to perform only limited error checking or not check results of tuned kernels at all. Otherwise, computing the full reference result would remove the performance benefit of the non-blocking tuning. \todo{try to reformulate it to be more clear and compact}

%SB? give a more meaningful subsection heading? (Example: Composite Tuning)
%JF it is showing also integration and dynamic tuning, I suggest to use more general heading, but not neccessary "Code Example" %SB ok
\subsubsection{Code Example}

Let us assume we have two kernels, \texttt{foo(a)} and \texttt{bar(b)}. The kernel \texttt{foo} produces a 2D array, which is used as an input for kernel \texttt{bar}: \texttt{b = foo(a); c = bar(b);}.
%\begin{lstlisting}
%b = foo(a)
%c = bar(b)
%\end{lstlisting}
Let us further assume a tuning parameter B\_TRANS, which determines if \texttt{b} is stored transposed. Clearly, the value of B\_TRANS must be the same for both \texttt{foo} and \texttt{bar}, so the kernels must be tuned together. Thus, we create a kernel composition with a tuning manipulator calling both kernels. The tuning manipulator is shown in Listing~\ref{lst:example_manipulator}. The class inherited from tuning manipulator must override the method \texttt{launchComputation}, which is responsible for executing the two kernels via KTT in our example, but it could also implement computation in C++ or call KTT functions for data movement or synchronization. 

\begin{footnotesize}
\begin{lstlisting}[caption={Tuning manipulator},mathescape,escapeinside={(*}{*)}, label=lst:example_manipulator]
class TunableFoobar : public ktt::TuningManipulator {
public:
  TunableFoobar(ktt::KernelId foo, ... ) :
    // assign kernels and input/output
    // to internal structures
  {}
  void launchComputation(const ktt::KernelId) 
  override {
    // tuning parameters can be queried here
    runKernel(foo);
    runKernel(bar);
  }
private:
  ktt::KernelId foo;
  ...
};
\end{lstlisting}
\end{footnotesize}

The code setting up KTT is sketched in Listing~\ref{lst:example_app}. It initializes the tuner at line 2, creates kernels (lines 5-8), their arguments (lines 11-12), and constructs a composition of the kernels (lines 16-22). The composition is created with a tuning manipulator implemented in a class \texttt{TunableFoobar} (see Listing~\ref{lst:example_manipulator}). The kernels are created with an initial configuration of NDRange and work-group size (lines 3-4), but this configuration can be altered in two ways: by defining the relation of NDRange/group size and some tuning parameter (using a pre-defined or lambda function), or directly in launchComputation method by any user code.

\begin{tiny}
\begin{lstlisting}[caption={Tuning initialization},mathescape,escapeinside={(*}{*)}, label=lst:example_app]
  // Initialize tuner and kernels foo, bar
  ktt::Tuner tuner(platformIndex, deviceIndex); 
  const ktt::DimensionVector ndRange(inputSize);
  const ktt::DimensionVector workGroup(128);
  ktt::KernelId foo = tuner.addKernelFromFile(
    kernelFile, "foo",  ndRange, workGroup);
  ktt::KernelId bar = tuner.addKernelFromFile(
    kernelFile, "bar", ndRange, workGroup);
    
  // Creation of kernel arguments a, b, c
  ktt::ArgumentId a = tuner->addArgumentVector(srcA, 
    ktt::ArgumentAccessType::ReadOnly);
  ...

  // Creation of composition and setting of arguments
  ktt::KernelId compositionId = tuner.addComposition(
    "foobar", std::vector<ktt::KernelId>{foo, bar}, 
    std::make_unique<TunableFoobar>(foo, bar, a, b, c));
  tuner.setCompositionKernelArguments(compositionId, 
    foo, std::vector<size_t>{a, b});
  tuner.setCompositionKernelArguments(compositionId, 
    bar, std::vector<size_t>{b, c});
    
  // Addition of tuning variables
  tuner.addParameter(compositionId, "B_TRANS", {0, 1});
\end{lstlisting}
\end{tiny}

After the setup, we can perform kernel tuning. Here, we demonstrate non-blocking dynamic autotuning, which is performed in the main application loop as sketched in Listing~\ref{lst:example_loop}. In our simple example, we use the variable \texttt{tuningOn} to specify whether dynamic tuning is performed (it can be set by some user-defined function to \textit{true} for a fixed number of iterations, or until some predefined performance is reached). The execution of a composition calling \texttt{foo} and \texttt{bar} can be achieved by two  methods: \texttt{runKernel} or \texttt{tu\-ne\-Ker\-nel\-By\-Step}. The \texttt{runKernel} executes the composition and stores result in variable \texttt{c}. The execution is performed with a tuning configuration defined by the programmer; usually, the fastest configuration is used. The second method, \texttt{tuneKernelByStep}, also performs the computation and stores results in \texttt{c}, but with a new values of the tuning parameters (selected by KTT using the selected search method). 
%SB? - are always all possible tuning configurations explored - what happens if only a subset is explored - then we cannot be sure that we get the fastest one - please clarify
%JF it should be clear now -%SB still the following sentence is a bit unclear - rewrite? %JF I have tried different formulation -- it is simple statement that when tuning space is already searched, there is no more search performed, but the fastest configuration is used, which is something one woul expect %SB better now
 If the tuning space has already been explored, the method \texttt{tuneKernelByStep} executes the configuration, which results in the fastest computation (so it behaves like \texttt{runKernel} executed with the best configuration). If the application is exploring only a subset of the tuning space, it can query the fastest known configuration via the \texttt{getBestComputationResult} method. The rest of the application does not need to be aware whether tuning is performed: the result \texttt{c} is obtained in any case.

\begin{tiny}
\begin{lstlisting}[caption={Main loop performing computation},mathescape,escapeinside={(*}{*)}, label=lst:example_loop]
  while(application_run) {
    ..
    if (tuningOn)
      tuner.tuneKernelByStep(compositionId, {c});
    else {
      ktt::ComputationResult best = 
        tuner->getBestComputationResult(compositionId);
      tuner.runKernel(compositionId, 
        best.getConfiguration(), {c});
    }
    // c is computed here
    ...
  }
\end{lstlisting}
\end{tiny}

\subsection{Independent Queues and Non-blocking Calls}
\label{sect:sync}

Accelerated codes often employ task-level parallelism to overlap computation on a host, computation on a device, and data movements between the host and the device. Moreover, simultaneous kernel execution may improve the performance of independent kernels when some kernels do not fully utilize the device. Task-level parallelism is realized via non-blocking kernel calls, asynchronous copy and also via multiple queues (OpenCL) or streams (CUDA).

In order to reach high performance when integrated into an application, KTT must support this functionality for the tuned kernels. Thus, it is possible to use queues (when using CUDA, KTT queues are implemented as CUDA streams) and non-blocking calls with KTT. However, during the tuning of the kernel, concurrent kernel execution or non-blocking execution may bias benchmarking (\eg{}, with concurrent kernel execution, the host code can execute another kernel at the device where the tuned kernel is running, so the measuted runtime of the tuned kernel increases). The bias in benchmarking could result in a wrong selection of the best tuning parameter values. Therefore, there are two types of task-level parallelism implemented in KTT:
\begin{itemize} % z popisu mi neni moc jasne jaky je mezi nimi rozdil. Pokud nekdo zavola kernel mimo KTT, tak to KTT prece stejne nemuze vyresit, ne? JF: KTT pocita s tim, ze se vse pousti pres nej (to piseme vyse), pokud mu do GPU matle dalsi kod, nic s tim nenadelame, ale nemelo by si v case mereni po GPU matlat necim jinym samo... je to treba jeste dovysvetlit v textu?
  \item \textit{intra-manipulator parallelism} allows simultaneous kernel execution and overlapping computations and memory copy inside a \texttt{launchComputation} method of a tuning manipulator;
  \item \textit{global parallelism} also allows simultaneous kernel execution and non-blocking kernel calls at the level of the application code, so the host code can call \texttt{runKernel} in non-blocking mode, allowing to overlap execution of multiple manipulators, or host and device computation.%outside of the tuning manipulator.
\end{itemize}

During the tuning process, global parallelism is not allowed, so only one tuning configuration is executed at a time. Therefore, benchmarking is not biased by executing another code on a computing device or in a CPU thread where KTT is running. However, tuning manipulators may still use intra-manipulator parallelism, so it is still possible to, \eg{}, execute multiple independent kernels in parallel, or overlap kernel execution with the data copy or the CPU code. 

When the tuning process ends, KTT also allows the global parallelism so that kernels or composition calls can be overlapped with another device or host code. Note that the result of the kernel or composition is downloaded to the host memory by default, which enforces synchronization. However, the user can create persistent arguments, which are not copied to the host by KTT unless the application explicitly calls the proper KTT copy method.

\subsection{Limitations}
\label{sect:limitations}

Recall that KTT forms an intermediate layer between a tuned application and the OpenCL or CUDA API. Therefore, it has to implement the interface to operate those APIs. The current implementation of KTT does not support all the features of CUDA and OpenCL. Due to the lack of OpenCL 2.0 implementation for NVIDIA GPUs, the OpenCL support is limited to OpenCL 1.2 with KTT. Also, some features of CUDA are not supported: texture, surface, and constant memory and cooperative grids. We believe that there is no fundamental problem to support those features in a future version of KTT.

The new features of CUDA and OpenCL, which require changes in kernel code only, do not require any explicit support in KTT, as KTT methods replace only the host API (for example, new warp-level synchronization or warp-matrix operations executed on new CUDA tensor cores can be used with KTT without any explicit support).

In its current implementation, a single instance of KTT works with a single computing device. To use multiple devices (\eg{}, in multi-GPU machine), the programmer has to create multiple instances of KTT and partition the tuning space manually. It also implies that there is no explicit support for tuning which device is to be used for which particular kernel.

\section{Autotuning Benchmarks}
\label{sect:benchmarks}

%SB? does the benchmark set just contain codes that have been tuned with KTT or the tunable codes as well? As I understood, each code in the benchmark set is a C++ code + KTT macros + CUDA/OpenCL kernels. If this is the case, this should be clearly explained in the paper. And, maybe, the term ``tunable'' benchmarks instead of tuned benchamrks should be used.
In this section, we introduce a set of ten tunable benchmarks. Each benchmark contains a C++ code, which prepares data and performs tuning with KTT, and OpenCL or CUDA code of tunable kernels. We briefly introduce their implementation and evaluate the benefits of autotuning by measuring their efficiency and assessing their performance portability. All benchmarks have been tuned for and
evaluated on seven different hardware devices as listed in Table~\ref{tab:hw}.
%
%\subsection{Testbed Setup}
%

\begin{table}[ht]
\centering
\small
\begin{tabular}{|l|l|r|r|}
\hline
Device                  & Architecture  & SP perf.  & BW \\ \hline
$2\times$ Xeon E5-2650 	& Sandy Bridge  & 512	    & 102\\
Xeon Phi 5110P    	& Knights Corner & 2,022    & 320\\
Tesla K20        	& Kepler        & 3,524     & 208\\
GeForce GTX 750  	& Maxwell       & 1,044     & 80 \\
GeForce GTX 1070 	& Pascal        & 5,783     & 256\\
Radeon RX Vega 56   	& GCN 5         & 8,286     & 410 \\
GeForce RTX 2080Ti 	& Turing      	& 11,750    & 616 \\
\hline
\end{tabular}
\caption{Devices used in our benchmarks. Arithmetic performance (SP perf.) is measured in single-precision GFlops, memory bandwidth (BW) is measured in GB/s.}
\label{tab:hw}
\end{table}

%\begin{table*}[ht]
%\centering
%\begin{tabular}{|l|l|r|r|r|}
%\hline
%Device                  & Architecture  & SP Performance    & Bandwidth & Released \\ \hline
%dual Intel Xeon E5-2650 & Sandy Bridge  & 512\,GFlops	    & 102\,GB/s & 2012\\
%Intel Xeon Phi 5110P    & Knights Corner & 2,022\,GFlops    & 320\,GB/s & 2012\\
%NVIDIA Tesla K20        & Kepler        & 3,524\,GFlops     & 208\,GB/s & 2012\\
%NVIDIA GeForce GTX 750  & Maxwell       & 1,044\,GFlops     & 80\,GB/s & 2014\\
%NVIDIA GeForce GTX 1070 & Pascal        & 5,783\,GFlops     & 256\,GB/s & 2016\\
%AMD Radeon RX Vega 56   & GCN 5         & 8,286\,GFlops     & 410\,GB/s & 2017\\
%NVIDIA GeForce RTX 2080Ti & Turing      & 11,750\,GFlops    & 616\,GB/s & 2018\\
%\hline
%\end{tabular}
%\caption{A list of devices used in our benchmarks.}
%\label{tab:hw}
%\end{table*}
%

\subsection{Tuning Parameters}
\label{sect:tuning_parameters}

With tuning of code optimization parameters, the tuning parameters can encode virtually any change of the source code. While many benchmarks contain tuning parameters performing the same type of optimization, their implementation may differ from case to case. In this section, we describe the common optimizations parameters implemented in most of the benchmarks.

\subsubsection{Work-group Size}
%SB? - what does ``reachable parallelism'' mean - degree of parallelism? granularity?
On GPUs, the size of work-group allows balancing the amount of reachable parallelism (\ie{}, amount of work-items which can run simultaneously) and allocated resources (\eg{}, private and local memory consumption). In general, smaller work-groups (to some extent) allow to allocate of more resources and reduce local barrier overhead. On the other hand, small work-groups may decrease memory locality when some type of memory blocking is used. Very small work-groups may also decrease reachable parallelism due to creation of under-populated warps or due to the limited amount of work-groups which can be placed on GPU simultaneously. On CPUs, work-items are processed in a vectorized loop and thus the work-group size mainly influences the amount of consumed registers and memory locality.

The optimization of work-group size (or block size in CUDA) is a common optimization method, which may be easily implemented without re-compilation of the kernel code. However, most of the integer arithmetic required for array indexing uses the work-group size. Consequently, when the work-group size is encoded by a tuning parameter, indexing arithmetics can be optimized during compilation.

\subsubsection{Work-item Coarsening}
Work-item coarsening (or thread coarsening in CUDA) is a well-known technique~\cite{volkov2008benchmarking, stone2007accelerating}, optimizing the amount of work per work-item. On GPUs, adding more work per work-item improves private memory locality and instruction-level parallelism. On the other hand, it also increases the number of used registers, so that the reachable parallelism can be reduced. Work-item coarsening is similar to the loop unrolling on CPUs, as each work-item (\ie{}, iteration of the generated vectorized loop) performs more computations.

\subsubsection{Caching in Local Memory}
Local memory (called shared memory in CUDA) is GPU-specific hardware, which allows work-items from the same work-group to share data. It is often used as an explicit cache, where data loaded from global memory are further processed (or where data are collected before they are moved to the global memory). Local memory is faster than global memory and usually also faster than global memory cache. On the other hand, explicit caching may be challenging with more complex memory access patterns. Therefore, it may or may not be efficient to cache data in local memory.

%SB? - not clear to me what ``local memory'' on CPU exactly refers to (cache? Local NUMA node?)
%JF - I think there is no exact definition where to put data in local memory in OpenCL standard -- it is always put in the global memory, but I have no idea if it is NUMA-aware (it probably depend on the OpenCL implementation). When local memory is reasonably small, it should be cached by standard caching mechanisms. - SB% OK
On CPUs, there is no special hardware for local memory -- data allocated in the local memory are placed in a buffer in the global memory. Therefore, there is no reason to use it for improving the speed of the code, but it can be still used to share data between work-items.

\subsubsection{Caching in Private Memory}
Private memory (or registers in CUDA) is the fastest memory available for both GPUs and CPUs. Explicit caching in private memory speeds-up access to the data. However, it may also lead to registers spilling on both GPU and CPU architectures.

\subsubsection{Tile Size}
Memory tiling is a common technique to improve spatial or temporal locality. It is usable for direct global memory access (a tile is stored in the cache by the hardware), or explicit caching in local or private memory. The tile size may or may not be equal to the work-group size (\eg{}, work-items can process multiple data elements, so the tile size is an integer multiple of work-group size). Bigger tiles ensure better cache locality as long as cache capacity is not exceeded. However, with explicit caching on GPUs, bigger tiles can reduce reachable parallelism by increasing resources consumption.

\subsubsection{Loop Unrolling}
Loop unrolling is a general technique, which allows increasing instruction-level parallelism, reducing branching and  simplifying array indexing by common subexpression elimination. It increases the performance of loops if there are enough registers available.

\subsubsection{Padding Local Memory}
GPU local memory consists of multiple banks (usually 32), which should be accessed in parallel to reach the highest performance. If different data from the same bank are read, a bank conflict occurs and the access into this bank is serialized, resulting in performance degradation. Padding arrays in  local memory can prevent bank conflicts in some situations. For example, parallel read of a column of a $32 \times 32$ matrix in  local memory results in a 32-way bank conflict. However, when the matrix is stored as $33 \times 32$ array, there is no conflict in accessing columns.

\subsubsection{Explicit Vectorization}
The code performed by work-items can be written in a vectorized form. Such a case is similar to loop unrolling with slightly modified effect. With GPUs, it is easier for the compiler to generate faster vector instructions for memory access (both global and local). With CPUs, the OpenCL compiler by default performs de-vectorization and vectorization, but it can be hinted to directly translate vectorized code into vector instructions, which can help if implicit vectorization is not efficient enough. On the other hand, explicit vectorization often increases register usage on GPUs. It may also increase the amount of workload per work-group, which increases registers pressure in case local barriers are called within the kernel.

\subsection{Benchmark Set Implementation}
Here, we introduce the implementation of the benchmark set used in this paper. As the development of autotuning benchmarks is quite a time consuming task (the tuning parameters have to be identified in the code, and their effect has to be implemented), we have composed a benchmark set from already available kernels, kernels developed by our group in several projects, and kernels developed as autotuned variants of previously available non-autotuned kernels. 
The benchmarks set covers important computational problems spanning across multiple application domains: image processing (3D Fourier Reconstruction and 2D Convolution), linear algebra (BiCG, GEMM, GEMM Batched, Matrix transpose, and Reduction), computational chemistry (Direct Coulomb Summation) and differential equation solvers (N-body and Hotspot). Most of the benchmarks use tuning parameters for performing the optimizations introduced in Section~\ref{sect:tuning_parameters}. Table~\ref{tab:benchmarks-parameters} shows which optimizations are implemented by which particular benchmark. Benchmarks which have been published previously are described briefly here, whereas the unpublished benchmarks are introduced in greater detail. Multiple benchmarks also implement special optimizations not listed in the table -- in such case, the optimizations are mentioned in the benchmark description in this section.

The benchmark set is publicly available. Except for 3D Fourier Reconstruction, all benchmarks are bundled with the Kernel Tuning Toolkit as examples of its usage\footnote{\url{https://github.com/Fillo7/KTT/examples}}. The autotuned version of 3D Fourier Reconstruction is currently not integrated into the production version of Xmipp, but it can be downloaded from Github\footnote{\url{https://github.com/I2PC/scipion/tree/jd_reconstructFourier_KTT}}.

\begin{table*}
\centering
\small
\begin{tabular}{|l|c|c|c|c|c|c|c|c|}
    \hline
    Benchmark 		& WG size 	& coarsening	& LM caching	& PM caching	& Tile size	& unrolling 	& LM padding 	& vectorization \\
    \hline
    BiCG		& \checkmark	& \checkmark	& \checkmark	& 		& \checkmark	& \checkmark	&  		& \\
    2D Convolution	& \checkmark	& \checkmark	& \checkmark	& \checkmark	& \checkmark	& \checkmark	& \checkmark	& \checkmark\\
    Coulomb 3D		& \checkmark	& \checkmark	&  		& 		& 		& \checkmark	&  		& \checkmark\\
    GEMM		& \checkmark	& \checkmark	& \checkmark	& 		& \checkmark	& \checkmark	& 		& \checkmark\\
    GEMM Batched	& \checkmark	& \checkmark	& \checkmark	& \checkmark	& 		& \checkmark	& \checkmark	& \checkmark\\
    Hotspot		& \checkmark	& \checkmark	& \checkmark	& 		& \checkmark	& \checkmark	&  		& \\
    Matrix Transpose	& \checkmark	& \checkmark	& \checkmark	& 		& \checkmark	& 		& \checkmark	& \checkmark\\
    N-body		& \checkmark	& \checkmark	& \checkmark	& \checkmark	& 		& \checkmark	&  		& \checkmark\\
    Reduction		& \checkmark	& \checkmark	&  		& 		& 		& 		&  		& \checkmark\\
    Fourier		& \checkmark	& \checkmark	& \checkmark	& 		& \checkmark	& 		&  		& \\
    \hline
\end{tabular}
\caption{Common optimizations tuned by benchmarks. Tile size is marked when it can be configured differently than work-group size. The abbreviations used in the names of the columns are as follows: "WG" is work-group, "LM" is local memory, "PM" is private memory.}
\label{tab:benchmarks-parameters}
\end{table*}

\subsubsection{BiCG}
BiCG is a kernel used in the biconjugate gradient method. It computes
\begin{equation}
q = Ap \\
s = A^Tr
\end{equation}
where $A$ is a matrix and $p, q, r, s$ are vectors. We have adopted the implementation from PolyBench/GPU~\cite{gray2012polybench} and implemented kernel fusion and cache tiling similarly to our previous work~\cite{filipovic2013optimizing}. %The resulting tuning space is listed in Table~\ref{tab:bicg-tuning}. The size of the tuning space is 5,122. 
In addition to the parameters listed in Table~\ref{tab:benchmarks-parameters}, we have created tuning parameters changing the following properties of the code:
\begin{itemize}
  \item whether BiCG is computed by the fused kernel (loading matrix $A$ only once), or by two separate kernels computing $Ap$ and $A^Tr$;
  \item the amount of work per work-group (it can iterate over multiple tiles, improving memory locality of output vectors);
  \item how the reduction of resulting vectors is performed (can be reduced in local memory or global memory);
  \item how reduction is implemented (using atomic operations, or finishing reduction in a separate kernel).
\end{itemize}
The implementation uses the tuning manipulator, as tuning parameters change the execution of kernels (\eg{}, when atomics are not used, an extra kernel is needed to finish computation of vectors $q ,s$).

%\begin{table*}[ht]
%\scriptsize
%\begin{center}
%    \begin{tabular}{|l|l|l|}%{|l|p{2cm}|p{8cm}|}
%    \hline
%    Parameter & Values & Description \\
%    \hline
%    FUSED & 0, 1 & use fused (1) or separate (0) kernels for matrix-vector product \\
%    BICG\_BATCH & 1, 2, 4, 8 & number of matrix elements processed by a work-item \\
%    USE\_SHARED\_MATRIX & 0, 1 &store matrix tile in local memory (1) or work directly with global memory (0)\\
%    USE\_SHARED\_VECTOR\_\{1,2\} & 0, 1 & store input vector $p$ or $r$ in local memory (1) or work with global memory (0) \\
%    USE\_SHARED\_REDUCTION\_\{1,2\} & 0, 1 & reduce vectors $q$ or $s$ in shared memory (1) or global memory (0) \\
%    ATOMICS & 0, 1 & use atomic operations (1) for reduction or reduce them by iterative kernel execution (0) \\
%    UNROLL\_BICG\_STEP & 0, 1 & let unrolling on compiler (0) or fully unroll inner loop (1) \\
%    ROWS\_PROCESSED & 128, 256, 512, 1024 & number of processed rows of matrix $A$ by a work-group \\
%    TILE & 16, 32, 64 & tile size \\
%    \hline
%    \end{tabular}
%\end{center}
%\caption{Tuning parameters in BiCG.}
%\label{tab:bicg-tuning}
%\end{table*}

\subsubsection{2D Convolution}
The 2D convolution example using $7 \times 7$ filter is adopted\footnote{Our code uses the same kernel and tuning space, but the application is modified to use KTT API.} from the CLTune project~\cite{nugteren2015cltune}. The special tuning parameters determine the way of handling shared boundaries of tiles.

\subsubsection{Direct Coulomb Summation}
The direct Coulomb summation precomputes the 3D spatial grid of electric charge around a molecule, used, \eg{}, in molecular docking~\cite{goodsell1996automated}. We have introduced the autotuned implementation in~\cite{filipovic2017autotuning}. Here, we evaluate a 3D version of the published algorithm. The algorithm tunes, besides those mentioned in Table~\ref{tab:benchmarks-parameters}, the following parameters:
\begin{itemize}
  \item whether input atoms are stored in global or in constant memory;
  \item whether input atoms are stored as a structure of arrays or as an array of structures.
\end{itemize}

\subsubsection{GEMM}
The generalized matrix-matrix multiply (GEMM) is a standard part of BLAS~\cite{dongarra1990set}. Its performance is critical for many applications. We have adopted an example from the CLTune project~\cite{nugteren2015cltune} with a complex tuning space containing 241,600 configurations. The large tuning space is mainly caused by applying optimizations listed in Table~\ref{tab:benchmarks-parameters} in multiple dimensions. Moreover, tuning parameters are provided for switching between continuous and strided access to the input matrices.

\subsubsection{GEMM Batched}
\label{sect:batched_gemm}
Regular BLAS implementations are optimized for large data vectors and matrices. However, some applications, such as deep learning~\cite{tensorflow2015whitepaper}, multifrontal solvers for sparse linear systems~\cite{duff1983multifrontal} or Finite Elements Method~\cite{ljungkvist2014matirx} require executing many instances of BLAS routines operating on very small matrices. Therefore, batched operations (\ie{}, grouping many BLAS calls that process small matrices together into a single call) are being developed to exploit contemporary highly-parallel hardware.

It has been shown that autotuning enables reaching near-peak performance for batched GEMM using very small matrices (up to $32 \times 32$ elements)~\cite{masliah2016high}. The implementation of batched GEMM has to be altered for different sizes of matrices~\cite{masliah2016high}. We have implemented the batched GEMM kernel from scratch. It is optimized for very small matrices similarly to~\cite{masliah2016high}, but also for highly rectangular small matrices. Note that for small matrices, GEMM is memory-bound (it does not expose high flop-to-word ratio). Therefore, optimization strategies are different than for GEMM optimized for larger matrices, resulting in a significantly smaller tuning space. Since our GEMM Batched benchmark is optimized for small matrices only, for bigger matrices, the original GEMM benchmark should be used.

Our implementation uses highly-configurable parallelism. For output matrix of size $m \times n$, a work-group of size $m \times y \times z$ is created, where $y, z$ are tuning parameters. Parameter $y$  defines work-item coarsening: it determines the number of work-items in the y-dimension which process one instance of matrix multiplication and hence the number of elements processed by each work-item. Parameter $z$ determines the number of matrix multiplication instances computed by a work-group. Caching in local memory is also implemented for the output matrix: it can be written into global memory directly, or multiple matrices can be arranged in local memory and written together (improves memory coalescing).

\subsubsection{Hotspot}
The Hotspot kernel, used for calculating a heat distribution on a 2D surface, is based on a kernel from the Rodinia benchmark suite~\cite{che2009rodinia}. It implements a 2D finite differences method, which can exploit temporal locality (as it is executed iteratively). We have implemented tuning parameters listed in Table~\ref{tab:benchmarks-parameters}, and parameter allowing to tune the number of steps performed in a kernel call (balances temporal locality against redundant computation).
%The tuning space consists of 480 configurations.

\subsubsection{Matrix Transpose}
We have implemented autotuning for a tiled matrix transposition sample from NVIDIA CUDA SDK 10.0. The tuning parameters additional to those defined in Table~\ref{tab:benchmarks-parameters} are as follows:
\begin{itemize}
  \item transposition of work-items (work-items in a warp may read rows and store columns, or read columns and store rows);
  \item explicit prefetching into the cache.
\end{itemize}

\subsubsection{N-body} % tenhle jsem psal ja? Jestli jo, tak asi neni z SDK10, bo to vyslo pozdeji JF: jo psal, zmenil jsem to na 9 (on se stejne imho nezmenil)
The computation of gravitational forces between $n$ bodies in space is based on the code sample from NVIDIA CUDA SDK 9.0. It computes a gravitational force between all pairs of bodies, and thus is a very compute-intensive benchmark. We have added the tuning parameters allowing tuning of how input bodies are stored (array of structure or structure of arrays) and also the optimizations defined in Table~\ref{tab:benchmarks-parameters}.

\subsubsection{Reduction} % dokaze KTT predat lambdu? Da se misto scitani pouzit jina funkce a pretunit? JF: ne uplne primo (musel by dat retezec kodu), ale je to dost zajimava myslenka, nekde bychom tohle mohli ukazat
The reduction benchmark computes the sum of all elements in an input vector. We have used the autotuned implementation from our previous work~\cite{filipovic2017autotuning}. There are two special optimizations affected by tuning parameters and not listed in Table~\ref{tab:benchmarks-parameters}:
\begin{itemize}
  \item whether the reduction is performed with at most one global barrier only by a fixed number of work-items, or iteratively by multiple kernels scaling with the size of the reduced vector;
  \item whether, with the fixed number of work-items, the final reduction is performed by extra kernel invocation, or by utilizing atomic operations.
\end{itemize}

%\subsubsection{Radix Sort}
%The radix sort example has been adapted from SHOC benchmark~\cite{danalis2010scalable}. It is composed of three kernels: reduce, top scan and bottom scan. We have added following tuning parameters, forming a space of 60 configurations:
%\begin{itemize}
%  \item work-group size;
%  \item vector size (input data can be packed into vectors of different sizes);
%  \item number of threads used by reduction and bottnom scan (only single work-group is used with top scan).
%\end{itemize}
%\todo{prostor neni zrejme sestaveny dobre, globalni a lokalni velikost by se mela tunit pro kazdy kernel zvlast, pokud to alg. dovoli}

\subsubsection{3D Fourier Reconstruction}
\label{sect:fourier_tuning_space}
One of the computationally demanding steps in the image reconstruction pipeline in cryo-electron microscopy is a 3D Fourier reconstruction~\cite{abrishami2015fast}: the process when 2D samples of arbitrary orientation are inserted into the 3D volume. We have used the autotuned implementation introduced in our previous work~\cite{strelak2019gpu}. This implementation can be tuned for specific hardware and also for specific samples resolution. In contrast to other benchmarks, 3D Fourier Reconstruction is implemented in CUDA and therefore can be evaluated on NVIDIA GPUs only. %The 3D Fourier reconstruction builds a 3D volume in the Fourier space from a set of the Fourier transforms of specimen's 2D projections. Essentially, the reconstruction algorithm inserts the Fourier transforms of the 2D images, each with multiple different rotations, into a 3D grid using an interpolation kernel.

%We have introduced the CUDA-based GPU acceleration of the 3D Fourier reconstruction in~\cite{strelak2018gpu}. However, as the paper is currently not public, we describe its tuning space here. The 430 tuning configurations allows to modify:
The tuning space allows several optimizations not listed in Table~\ref{tab:benchmarks-parameters}:
\begin{itemize}
  \item atomic writing into output volume (allows to process multiple 2D samples in parallel);
%  \item number of 2D images processed in parallel;
  \item precomputation or on-the-fly computation of interpolation weights;
%  \item caching precomputed weights in shared memory or direct read from global memory.
%  \item storing image tile in shared memory or direct read from global memory;
  \item how are work-items mapped to data inside tiles of input 2D samples (optimizing cache locality).
\end{itemize}

\subsubsection{Summary}

Our benchmarks use a variety of tuning parameters, some of them common for multiple benchmarks, some of them specific for a given computational problem. The size and dimensionality of tuning spaces are summarized in Table~\ref{tab:benchmarks-spaces}. Note that the number of tuning parameters can be higher than the number of tuned optimizations described in this section, because some optimizations are implemented by multiple tuning parameters (\eg{}, if optimizations are applied to multiple buffers or multiple dimensions independently). Several benchmarks have been executed with a smaller tuning space on Radeon Vega56 because the AMD ROCm driver has been crashing with some tuning configurations (mainly using vectors of size 16 and higher loop unrolling factors). Those benchmarks are Direct Coulomb Summation, GEMM, and N-Body.

The tuning spaces of benchmarks have been defined during their development. We have not performed any \textit{a posterior} adjustment of the tuning spaces based on the experimental evaluation (\eg{}, removing poorly-performing configurations). Therefore, we are able to evaluate the difficulty of searching tuning space without bias caused by experimental knowledge of well- or poor-performing configurations.

\begin{table}
\centering
\small
\begin{tabular}{|l|r|r|}
    \hline
    Benchmark 	& dimensions & configurations \\
    \hline
    BiCG	& 11 & 5,122 \\
    Convolution	& 10 & 5,248 \\
    Coulomb 3D  & 8 & 1,260 \\
    GEMM	& 15 & 241,600 \\
    GEMM batched & 11 & 424 \\
    Hotspot	& 6 & 480 \\
    Transpose   & 9 & 10,752 \\
    N-body	& 8 & 9,408 \\
    Reduction	& 5 & 175 \\
    Fourier 	& 6 & 360 \\
    \hline
\end{tabular}
\caption{A list of the benchmarks and the size and dimensionality (\ie{}, the number of tuning parameters) of their tuning spaces.}
\label{tab:benchmarks-spaces}
\end{table}

\subsection{Efficiency of Benchmarks}

\begin{table}
\centering
\small
\begin{tabular}{|l|l|l|p{3cm}|}
    \hline
    Benchmark 	& bound. 	& ops.	& note \\
    \hline
    BiCG	& mem		& $4 a^2$ 	& $a$: width and height of the input matrix \\
    Coulomb 3D  & comp		& $6 a k^3$	& $a$: number of atoms, k: number of grid points per dimension \\
    GEMM	& comp		& $2 a^3$	& $a$: width and height of all matrices \\
    GEMM batched & mem		& $12 n a^2$	& $a$: width and height of all matrices, $n$: number of matrices\\
    Hotspot	& mem	  	& $4 i a^2$	& $a$: width and height of the input matrix, $i$: number of iterations\\
    Transpose   & mem		& $8 a^2$	& $a$: width and height of all matrices \\
    N-body	& comp		& $20 n^2$	& $n$: number of bodies\\
    Reduction	& mem		& $4 n$		& $n$: size of input vector\\ % uplne nevidim kde se bere ta '4' JF: viz popisek tabulky, je to v bytech
%    Sort 	& memory	&		& \\
    \hline
\end{tabular}
\caption{Number of operations performed by different benchmarks. The column "bound." 
 distinguishes between memory-bound codes (operations in "ops." column refer to transferred bytes) and compute-bound codes (operations in "ops." column refer to flops).}
\label{tab:benchmarks-flops}
\end{table}

%\begin{table*}
%\centering
%\small
%\begin{tabular}{|l|l|l|l|}
%    \hline
%    Benchmark 	& boundary 	& operations	& note \\
%    \hline
%    BiCG	& memory	& $4 a^2$ 	& $a$: width and heigh of the input matrix \\
%    Coulomb 3D  & compute	& $6 a k^3$	& $a$: number of atoms, k: number of grid points per dimension \\
%    GEMM	& compute	& $2 a^3$	& $a$: width and heigh of all matrices \\
%    GEMM batched & memory	& $12 n a^2$	& $a$: width and heigh of all matrices, $n$: number of matrices\\
%    Hotspot	& memory  	& $4 i a^2$	& $a$: width and heigh of the input matrix, $i$: number of iterations\\
%    Transpose   & memory	& $8 a^2$	& $a$: width and heigh of all matrices \\
%    N-body	& compute	& $20 n^2$	& $n$: number of bodies\\
%    Reduction	& memory	& $4 n$		& $n$: size of input vector\\ 
%%    Sort 	& memory	&		& \\
%    \hline
%\end{tabular}
%\caption{Performance boundaries of benchmarks.}
%\label{tab:benchmarks-flops}
%\end{table*}

If we want to study autotuning spaces (especially concerning how hard it is to search them), we should first prove that those spaces allow us to generate a code with high performance. Here, we demonstrate that our benchmarks either reach performance close enough to theoretical boundaries of the hardware or at least outperform the baseline\footnote{the implementation we used as a basis for our benchmark, \eg{}, the Rodinia's Hotspot} implementation significantly. We do not evaluate 2D Convolution here: it does not perform at peak performance, but it reaches state-of-the-art performance~\cite{nugteren2015cltune}, so it can be considered efficient. 
%SB? next sentence unclear to me
%JF is this better? We simply cannot evaluate performance boundaries, because this code has chaotic memory access, so it waits for memory or cache latency for most of the time. However, this is stil much faster than previously published results, as we've shown in \cite{strelak2019gpu} %SB better now
We also exclude 3D Fourier Reconstruction -- it is a memory latency-bound code, making theoretical performance boundaries difficult to evaluate.
However, it has been shown that the autotuned implementation of our gather-based 3D Fourier Reconstruction significantly outperforms state-of-the-art scatter-based approach~\cite{strelak2019gpu}.

We define the efficiency of a benchmark as the relative performance of the benchmark with respect to the relevant hardware performance boundaries (memory or arithmetic throughput). More precisely, we use:
\begin{equation}
  \mathit{efficiency} = 100 \cdot \max(\frac{\frac{\mathit{MEMops}}{\mathit{time}}}{\mathit{MEMpeak}}, \frac{\frac{\mathit{ALUops}}{\mathit{time}}}{\mathit{ALUpeak}})
\end{equation}
where $\mathit{time}$ is the runtime of computation, $\mathit{MEMpeak}$ and $\mathit{ALUpeak}$ is peak memory and arithmetic throughput of the hardware\footnote{Only half the memory bandwidth has been considered for dual Intel Xeon E5-2650 because OpenCL provides no mechanism to optimize for NUMA in the dual-socket system (pinning memory buffers and work-groups to NUMA nodes is not possible). Therefore the full system bandwidth is not available.}. The $\mathit{MEMops}$ and $\mathit{ALUops}$ are the number of memory or arithmetic operations which are \textit{essential} to solve the task. % without any overhead. 
%(\ie{}, we consider operations which are always needed to compute the result and exclude operations which are replicated among work-items, used, \eg{}, for array indexing, communication, or duplicated computations). 
In other words, we count the number of operations required to solve the problem, not the number of operations required to execute the algorithm (such as array indexing, communication or computations duplicated among work-items). For example, BiCG benchmark is a memory-bound code, which essentially needs to read the input matrix A once. Therefore, even if the unfused implementation reads it twice, the number of operations is computed as the size of the input matrix in bytes divided by the runtime of the implementation.
The formulas for computing $\mathit{ALUflops}$ or $\mathit{MEMops}$ of the benchmarks are given in Table~\ref{tab:benchmarks-flops}. %Note that the BiCG, Matrix Transpose, Hotspot, Reduction and Radix Sort and Batched GEMM are memory-bound codes, whereas 3D Coulomb Sum, GEMM and N-body are compute-bound. 
%Note that the performance of the benchmark is computed as 

For all benchmarks, we have measured the performance with sufficiently large data as to fully utilize the GPUs. For Batched GEMM,  small matrices of size $16 \times 16$ have been used.

\begin{table*}
\centering
\small
\begin{tabular}{|l|l|l|l|l|l|l|l|}
    \hline
    Benchmark 	& 2080Ti 	& 1070 		& 750 		& K20 		& Vega56	& E5-2650 	& 5110P \\
    \hline
    BiCG	& 88.3\% 	& 84.7\% 	& 81.7\% 	& 50.4\%	& 75.6\%	& 46.0\%	& 6.45\% \\
    Coulomb 3D  & 91.8\%	& 91.4\%	& 84.3\%	& 43.2\%	& 65.3\%	& 74.2\%	& 22.2\% \\
    GEMM	& 79.8\%	& 80.6\%	& 91.1\%	& 51.3\%	& 96.3\%	& 37.5\%	& 19.7\% \\
    GEMM batched &  86.8\%	& 81.4\%	& 90.0\%	& 49.6\%	& 86.0\%	& 27.7\%	& 20.9\% \\
    Transpose   & 87.1\%	& 80.2\%	& 86.3\%	& 64.2\%	& 86.1\%	& 62.5\%	& 10.0\% \\
    N-body	& 89.7\%	& 86.6\%	& 87.7\%	& 40.6\%	& 82.2\%	& 77.7\%	& 29.9\% \\
    Reduction	& 68.7\%	& 87.5\%	& 89.4\%	& 64.1\%	& 71.6\%	& 33.9\%	& 10.1\%\\ 
    \hline
    Hotspot	& $1.35\times$  & $1.94\times$  & $2.06\times$  & $1.4\times$   & $2.88\times$  & $1.2\times$   & $12.8\times$ \\ % hotspot bych mozna oddelil nejakou tlustci carou. Kdyz jsem cetl tu tabulku, tak se mi mixovaly procenta a nasobky. JF lepsi?
%    Sort 	&		&		&		&		&		&		& \\
    \hline
\end{tabular}
\caption{Performance of benchmarks autotuned for various hardware devices. The performance relative to the theoretical peak of devices (see Table~\ref{tab:benchmarks-flops}) is shown for all benchmarks except for Hotspot, which is compared to the baseline Rodinia implementation.}% \todo{maybe put bar figure instead of graph here, but Hotspot must be solved somehow}}
\label{tab:benchmarks-perf}
\end{table*}

The efficiency of tuned implementations is given in Table~\ref{tab:benchmarks-perf}. The performance of Hotspot benchmark is not close to the theoretical peak, so we measure the speedup over Rodinia implementation. The number of steps per kernel invocation is exposed to the user as a parameter in Rodinia's implementation of Hotspot. To have a fair comparison, we have searched for the best-performing number of steps manually, testing the same values which have been tested by KTT in the autotuned version. As we can see, the performance on GPUs is very good in general. We can reach a performance close to the theoretical peak (75\% or more) in most cases for all architectures except Kepler (Tesla K20), which is less efficient than other architectures in all benchmarks. The performance on dual-CPU (Xeon E5-2650) and MIC (Xeon Phi 5110P) is often far from the theoretical peak. The development of OpenCL compiler seems to be not of high priority for CPU-based systems (for example Xeon Phi is not supported in Intel OpenCL from 2015), so this result is not surprising. 

Note that the performance of Coulomb 3D and N-body benchmarks has been computed differently for GeForce GTX 2080Ti: the Touring architecture seems to perform transcendental functions in parallel to FP32 instructions. Therefore, we have excluded the reciprocal square root from the computation of overall floating-point operations, using formulas $5 a k^3$ and $19 n^2$ for Coulomb 3D and N-body, respectively (see Table~\ref{tab:benchmarks-flops}). Otherwise, the performance would be overestimated.% -- the Coulomb 3D benchmark would perform at more than 100\%.

\subsection{Performance Portability}

In this section, we evaluate the performance portability of benchmarks without re-tuning them -- \ie{}, how benchmarks perform if they are executed on a different device than they are tuned for. This evaluation has been performed as follows. We have tuned all benchmarks for all devices $d$. Then, for each benchmark tuned for device $d_i$, we have measured its performance on devices $d_j, j \neq i$. The performance is computed as a percentage of the maximal reachable performance for the device: $100\frac{\mathit{perf}(d_j)}{\mathit{perf}(d_i)}$. 

Let us compute the performance portability between GeForce GTX 750 and GeForce RTX 2080Ti as an example. When BiCG benchmark is tuned for GeForce GTX 750, it reaches 65\,GB/s, when tuned for GeForce RTX 2080Ti, it reaches 544\,GB/s. When the code tuned for GeForce RTX 2080Ti is executed on GeForce GTX 750, it reaches 54.6\,GB/s, so the performance portability is 84\%. When the code tuned for GeForce GTX 750 is executed on GeForce 2080Ti, it reaches performance 381\,GB/s, so the performance portability is 70\%.

Due to vast number of combinations, we compact the results in Table~\ref{tab:benchmarks-portability} in the following way: we show the average with standard deviation and worst-case performances (i) across GPU architectures, (ii) when GPU code is executed on CPU-derived architecture (CPU and MIC) and (iii) when CPU or MIC code is executed on GPU architecture.

\begin{table*}
\centering
\small
\begin{tabular}{|l|lll|lll|lll|}
    \hline
    	 	& \multicolumn{3}{c|}{GPU$\rightarrow$GPU}	& \multicolumn{3}{c|}{GPU$\rightarrow$CPU/MIC}	& \multicolumn{3}{c|}{CPU/MIC$\rightarrow$GPU} \\
    Benchmark	& avg$\pm$stdev		& worst	& failed		& avg$\pm$stdev		& worst	& failed		& avg$\pm$stdev		& worst	& failed \\
    \hline
    BiCG	& 89.0\%$\pm$12.3\%	& 57\% 	& 1		& 44.1\%$\pm$17\%	& 28\% 	& 0		& 38.8\%$\pm$29.5\%	& 11\% 	& 0 \\ %checked
    Convolution	& 79.4\%$\pm$14.9\%	& 55\%	& 3		& 56.9\%$\pm$18.5\%	& 33\%	& 0		& 10.0\%$\pm$3.6\%	& 6\% 	& 1 \\ %checked
    Coulomb 3D  & 95.8\%$\pm$6.5\%	& 67\%	& 0		& 84.8\%$\pm$2.7\%	& 81\%	& 0		& 23.3\%$\pm$16.9\%	& 3\%	& 2 \\ %checked
    GEMM	& 83.6\%$\pm$16.4\%	& 31\% 	& 0		& 18.6\%$\pm$18.5\%	& 1\%	& 0	 	& 22.3\%$\pm$6.6\%	& 13\% 	& 2 \\ %checked
    GEMM batched & 85.4\%$\pm$17\%	& 37\%	& 0		& 68.2\%$\pm$13.2\%	& 39\%	& 0		& 76.7\%$\pm$22.2\%	& 46\%	& 1 \\ %checked
    Hotspot	& 80.3\%$\pm$17.5\%	& 46\%	& 3		& 70.3\%$\pm$15.6\%	& 44\%	& 0		& 65.1\%$\pm$8.9\%	& 59\%	& 6 \\ %checked
    Transpose   & 85.0\%$\pm$21.9\%	& 8\%	& 3		& 51.0\%$\pm$27.1\%	& 11\%	& 0		& 34.7\%$\pm$14.7\%	& 14\%	& 0 \\ %checked
    N-body	& 78.8\%$\pm$24.2\%	& 2\%	& 3		& 45.9\%$\pm$30.1\%	& 0\% 	& 0		& 25.7\%$\pm$15.6\%	& 6\%	& 2 \\ %checked
    Reduction	& 88.4\%$\pm$24\%	& 12\%	& 3		& 53.1\%$\pm$17.4\%	& 26\%	& 0		& 68.3\%$\pm$23.8\%	& 37\%	& 1 \\ %checked
    Fourier 	& 74.5\%$\pm$30\%	& 31\%	& 0 		& N/A			& N/A	& N/A		& N/A 			& N/A	& N/A \\
    \hline
\end{tabular}
%SB -changed N/A to failed in the respective column heading to differentia from the N/A used for Fourier
\caption{Relative performance of benchmarks ported across GPU architectures, from CPU/MIC to GPU and from GPU to CPU/MIC, without re-tuning. Avg$\pm$stdev denotes the average and standard deviation of the relative performance, worst shows the worst-case performance, and failed shows the number of cases when some configuration cannot be executed on a device. 3D Fourier Reconstruction has been executed on samples of $128 \times 128$ pixels on NVIDIA GPUs except for K20.}
\label{tab:benchmarks-portability}
\end{table*}

The table clearly demonstrates that the performance is not portable in general. Although the average performance portability is not bad among GPUs, the worst-cases are showing that for some benchmarks, there are combinations of GPUs with very bad performance portability, suggesting the autotuning should be re-executed for different architectures. This is in line with related work, such as~\cite{kurzak2012autotuning, gonzalo2017revisiting, strelak2019gpu}. The performance portability is much worse in the case when the benchmarks are tuned for a CPU or MIC and executed on a GPU and vice versa. The poor portability between GPUs and CPU or MIC emphasizes the important role of tuning -- although our benchmarks cannot reach peak performance on CPU or MIC, their performance is much higher than in case when the GPU-tuned code is simply executed on a CPU or MIC.

This experiment also reveals a serious limitation of functional portability with OpenCL. OpenCL guarantees the functional portability of the code if it can be executed on a device. When a kernel uses more hardware resources (\eg{}, number of registers per work-item) than is available on the device, it cannot be executed. It seems that this is the case of finely-tuned kernels, which often use as many resources as possible. When such kernels are executed on a device with a lower amount of resources, they fail. As it is shown in Table~\ref{tab:benchmarks-portability}, this can happen when a code tuned for CPU, MIC, or GPU is executed on a different GPU.

\section{Dynamic Autotuning}
\label{sect:dynamic_eval}
In this section, we experimentally evaluate dynamic autotuning for two applications: batched GEMM and 3D Fourier reconstruction. Moreover, we analytically determine the potential of dynamic autotuning for the rest of the benchmarks.

\subsection{Methodology}
%SB - I susggest to recap here the main charactersitics of dynamic tuning from Section 3.2
%JF better?  %SB should be oky now, commented stuff in () out
Recall that with dynamic autotuning, the tuning space is explored at runtime during application execution.
%SB comments this out - becuase ``after'' sounds like after the application has terminate (\eg{}, after its first execution on a given hardware, or after input data change). - 
Therefore, the implementation variants are compiled and benchmarked during application run-time, resulting in four sources of overhead:
\begin{itemize}
    \item compilation of OpenCL or CUDA kernels (each explored tuning configuration needs to be compiled by the JIT compiler); 
    \item execution of slower kernels (slower kernels prolong tuning time even in non-blocking autotuning when their results are used for computation);
    \item enforced global synchronization between tuning runs (during autotuning, execution of the tuning manipulators is not overlapped, see Section~\ref{sect:sync});
    \item testing kernel output (this step is optional).
\end{itemize}
These overheads are relevant during the tuning phase only (\ie{}, when new configurations are searched). However,  when a sufficient number of tuned kernel invocations is performed after the tuning, the overhead becomes negligible. Here, we want to know how long the dynamically-tuned code has to run to amortize the tuning overhead under a certain value. Or, alternatively, if dynamic autotuning reaches better performance than a code that has been offline-tuned for a different device or input data.

Note that the overhead of the KTT API (mainly the execution of manipulator in function runKernel) is negligible during the execution of the tuned code. %Naturally, in some special cases, there may be some overhead introduced by the lack of some functionality in the current version of KTT. For example, if the code without autotuning uses dynamic parallelism, the dynamic parallelism must be removed when using KTT (see Section~\ref{sect:limitations}).

\subsection{Batched GEMM}
%SB? clarify that the previous bacthed GEMM uses matrices of fixed size (e.g., 1000 32x32 matrices) %JF done
%SB as I understand, for dynamic tuning a different variant of Batched GEMM has been developed where the size of matrices changes (at runtime) e.g., 100 32*32 matrices; 100 8*8 matrices, 100 16X16nmatrices .. JF: we use the same kernel, but different C++ code, which is changing matrix sizes and triggers tuning repeatedly -- it should be more clear in the text (subsubsection Implementation)
%SB maybe this should be somehow mentioned already here
%SB provide a brief motivation and mentione why dynamic tuning is necessary for Batched GEMM
%JF I tried to improve modivation in the paragraph below - %SB reads better now
%SB it is unclear to me how an offlined-tuned version for Batched GEMM with dynamically varying matrix sizes would work - please clarify
%JF it is explicitly mentioned in Implementation subsection now -%SB OK
In the previous section, we have introduced an autotuned kernel for batched multiplication of very small matrices. This kernel is tuned for fixed sizes of matrices, \eg{}, we have used matrices of size $16 \times 16$ for experiments in Section~\ref{sect:benchmarks}. However, the space of the possible matrix sizes is large. The GEMM kernel performs $C = A \cdot B$, where $A$ is an $i \times j$ matrix, $B$ a $k \times i$ matrix, and $C$ a $k \times j$ matrix. Considering small matrices of sizes up to 32 in each dimension $i,j,k$, we get 32,768 combinations of the sizes. Consider an application or library, which does not know the sizes of multiplied matrices before it is executed. It would be impractical to offline tune the application or library for all possible sizes, so dynamic tuning, performed at runtime once the matrix size is fixed, is of high practical value.

%To demonstrate performance portability, we have set a following experiment. The 1,000 random sizes $(a, b, c)$ where $a, b, c \in <2, 32>$ has been generated. For each random tuple, we have executed a batched matrix multiplication, computing $C = A B$, where $A$ is matrix of $a \times b$ size, $B$ is matrix of $c \times a$ size and $C$ is matrix of $c \times b$ size. For each matrix multiplication except the first one, we have measured efficiency of a tuned implementation and efficiency of implementation using optimal parameters from previous run, if executable (\ie{} demonstrating performance portability). The measured efficiency is shown in Table~\ref{tab:gemm_eff}. The table clearly demonstrates, that our implementation of batched GEMM is efficient, but tuning parameters needs to be adjusted when input size changes.

\subsubsection{Implementation}
%SB needs improvment; - are the results of the explored configurations used or not?
%JF clarified in the text -%SB OK
We have prepared an experiment, which simulates a real application changing the matrix size from time to time. Our testing application\footnote{\url{https://github.com/Fillo7/KTT/blob/master/examples/gemm_batch/demo.cpp}} executed the tunable implementation of batched GEMM introduced in Section~\ref{sect:batched_gemm}, but it periodically changes the size of matrices and performs dynamic tuning. More precisely, the application computes batched GEMM in a loop and randomly changes sizes $i, j, k \in \interval{2}{32}$ every 30 seconds. The application does not save the results of dynamic autotuning, so every time the new sizes are used, the autotuning starts from scratch. The batch size has been selected so that matrices occupy approximately 900\,MB of memory (so enough parallelism is also exploited with very small matrices). When the size of matrices is changed, the dynamic tuning using random search starts and is performed until (i) a configuration reaching 75\,\% of the peak memory bandwidth is found, or (ii) 20 configurations have been explored. The first rule allows the application to stop tuning when a configuration resulting in a sufficient performance is reached (we call such a configuration as \textit{well-performing configuration}). The purpose of the second rule is to stop tuning when a configuration performing close to the theoretical peak cannot be easily found (or does not exist at all). After the tuning is stopped, the computation with the fastest tuning configuration continues until the sizes of matrices are changed again. With an application changing matrices sizes less often, the tuning time could be prolonged. We have measured the performance without tuning overhead (showing whether efficient kernels can be found under limited tuning budget) and with tuning overhead (showing the real performance of the dynamically tunned application). The time required for initialization and copying of newly created matrices was not benchmarked.

\subsubsection{Evaluation}

\begin{table}[ht]
\centering
\small
\begin{tabular}{|l|l|l|l|}
    \hline
    Device      & Maximum  	& Restricted 	& Incl. overhead \\ \hline
    E5-2650 	& 24.5\,GB/s	& 88.6\%	& 82.9\%	\\ %exhaustive full 24.5, limited budget exl. overhead 21.7, incl. overhead 20.3
    5110P    	& 22.9\,GB/s	& 82.1\%	& 72.1\%	\\ %exhaustive full 22.9, limited budget exl. overhead 18.8, incl. overhead 16.5
    K20         & 91.2\,GB/s	& 92.7\%	& 61.3\%	\\ %exhaustive 91.3, exhaustive full 91.2, limited budget exl. overhead 84.5, incl. overhead 55.9
    GTX 750   	& 63.0\,GB/s	& 91.4\%	& 87.8\%	\\ %exhaustive 61.1, exhaustive full 63.0, limited budget exl. overhead 57.6, incl. overhead 55.3
    GTX 1070  	& 205.7\,GB/s	& 97.2\%	& 94.3\%	\\ %exhaustive 201.2, exhaustive full 205.7, limited budget exl. overhead  200.0, incl. overhead 194.0
    Vega 56 	& 308.5\,GB/s	& 86.2\%	& 74.4\%	\\ %exhaustive full 308.5, limited budget exl. overhead XXX (requires more precise measurement), incl. overhead 229.6
    2080Ti 	& 523.4\,GB/s	& 92.6\%	& 85.3\%	\\ %exhaustive 496.1, exhaustive full 523.4, limited budget exl. overhead 484.5, incl. overhead 446.7
    \hline
\end{tabular}

\caption{Dynamically tuned Batched GEMM on different computational devices. The second column (Maximum) shows the average performance of the fastest configurations  (average for all tested matrix sizes). The third column (Restricted) shows averaged relative performance (relative to the maximum) of configurations reachable with dynamic tuning under limited tuning budget (at most 20 configurations explored). 
%SB relative performance? relative to what? 
%JF hopefully more clear now -%SB OK
Finally, the fourth column (Incl. overhead) shows the averaged relative performance of dynamically tuned code, including tuning overhead.}
%SB? dynamicylly tuned in this case means including the overhead of restricted search? %JF yes, it is complete time of search, compilation, tuning runs and execution of completely tuned code -- should it be mentioned somehow explicitly?
\label{tab:gemm-perf}
\end{table}
%SB needs improvment; would it make sense to include comparison to offline tuned code as well?
%JF it's there -- I hope it is better explained now - %SB OK
%SB ``measured 3000 seconds''  reformulated - please check %JF OK
We have run the experiments for 3,000 seconds (\ie{} 100 changes of matrix sizes) with all devices used in this paper. The matrix sizes have been selected randomly, but the same sizes are used for all devices. We have also performed offline tuning with exhaustive search for those sizes to obtain performance of the fastest configuration at each device. The performance with and without tuning overhead is computed as the relative performance of the fastest configuration found by the offline tuning. The results averaged over all matrix sizes are shown in Table~\ref{tab:gemm-perf}. The code executing on dual Xeon E5-2650, Xeon Phi 5110P and Tesla K20, is compiled on Xeon E5-2650, which has quite poor single-core performance and therefore requires a longer time for compilation. The prolonged compilation time does not limit performance on CPU and MIC significantly, because average kernels' runtime is high and therefore, the compilation does not induce significant overhead. On the other hand, the compilation overhead is quite noticeable with Tesla K20. The batched GEMM kernel performs in general very well (\ie{}, it is close to the theoretical peak) on GPUs except for Tesla K20. Its performance with overhead is also quite close to the peak kernel performance (85\% or better) in case of GeForce GTX 750, GeForce GTX 1070 and GeForce RTX 2080Ti, using Core i7-8700 for compilation. A bigger gap between 
%SB? peak kernel performance ?
%JF changed -%SB OK
the performance of the fastest kernels discovered during the tuning and performance with overhead can be seen with Radeon RTX Vega 56 (running in a system with Ryzen 7 1700). The main reason is the higher number of poorly performing configurations and, therefore, more tuning steps required to find 
%SB I think it is not compeletely celar what well-performing actually means here %JF hopefully better %SB yes
a well-performing configuration (\ie{}, the configuration within 75\% of peak memory bandwidth, which leads to finalization of the tuning). %SB? why is a higher number of steps required? %JF fixed

\begin{figure}[t]
    \centering
    \includegraphics[width=.9\hsize]{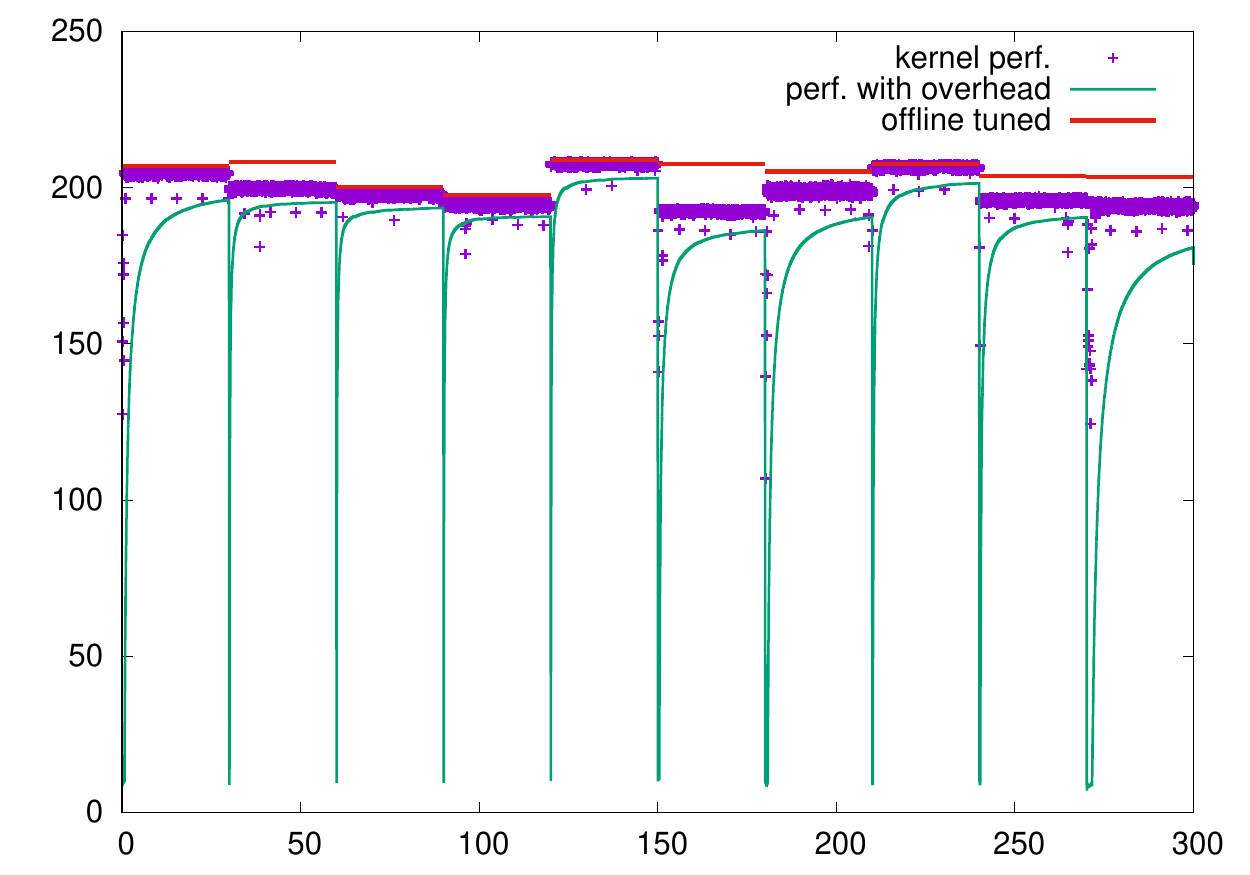}
    \caption{Performance of dynamically tuned batched GEMM on GeForce GTX1070 + Core i7-8700. The sizes of matrices are changed every 30\,s. Performance of actually executed kernels is depicted as dots, whereas lines show performance including overhead. The maximal performance reachable via offline tuning with exhaustive search is shown as horizontal red lines.}
    \label{fig:gemm-perf-1070}
\end{figure}

\begin{figure}[t]
    \centering
    \includegraphics[width=.9\hsize]{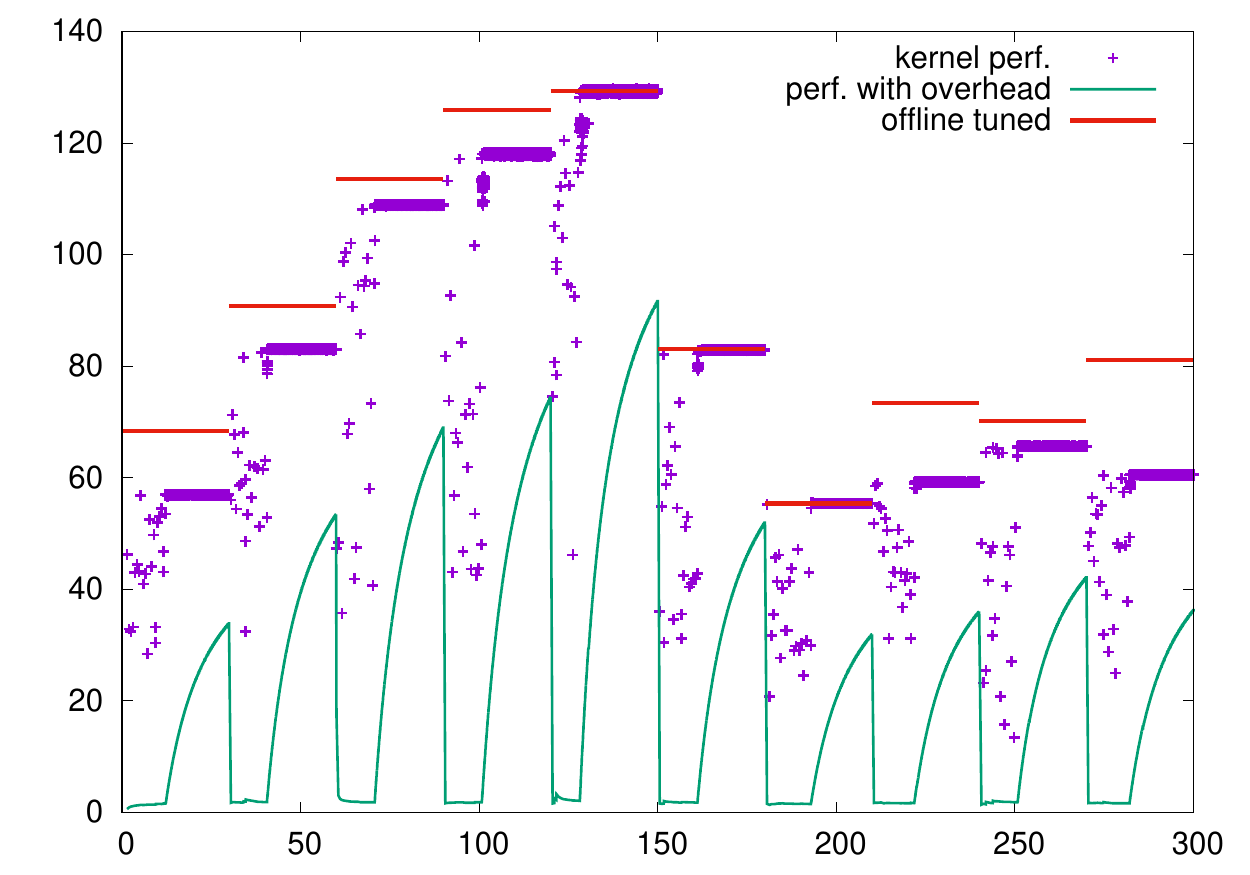}
    \caption{Performance of dynamically tuned batched GEMM on Tesla K20 + Xeon E5-2650. The sizes of matrices are changed every 30\,s. Performance of actually executed kernels is depicted as dots, whereas lines show performance including overhead.  The maximal performance reachable via offline tuning with exhaustive search is shown as horizontal red lines.}
    \label{fig:gemm-perf-k20}
\end{figure}

For better illustration of tuning performance, the first 300\,s of the benchmark execution is shown for well-per\-forming GeForce GTX 1070 in Figure~\ref{fig:gemm-perf-1070} and for Tesla K20, which suffers from tuning overhead, in Figure~\ref{fig:gemm-perf-k20}. Naturally, the performance including tuning overhead drops, when a new matrix size is used and increases in time as tuning overhead is amortized. It can be seen that GeForce GTX 1070, coupled with the modern processor Core i7-8700, is capable of amortizing tuning overhead in a very short time -- after 30 seconds of execution, performance with tuning overhead is close to peak kernel performance. Even if multiple configurations are searched before 
%SB? what is actaully a well-performing one? (some fraction of peak, e.g. 80%?) - was this clearly defined somewhere? maybe repeat here again %JF fixed above
a well-performing one 
is found (see performance between 270 and 300 seconds for GeForce GTX 1070 in Figure~\ref{fig:gemm-perf-1070}), the performance with tuning overhead is close to the performance of the best kernel found during the autotuning. On the other hand, Tesla K20 cannot reach high performance for many matrix sizes. This also means that more configurations are searched during the autotuning process. The system with K20 uses an older Xeon E5-2650, which also prolongs kernel compilation time. Therefore, the overhead of the tuning process is significant and would need more kernel invocations to amortize. Tesla K20 is also not able to reach high performance under limited tuning budget (see the difference between kernel performance and performance reachable by offline tuning between 210\,s and 240\,s, or between 270\,s and 300\,s for example).

\subsection{3D Fourier reconstruction in Xmipp}

We have introduced the CUDA-based GPU acceleration of the 3D Fourier reconstruction in~\cite{strelak2019gpu}, where KTT has been integrated into the 3D Fourier reconstruction for offline tuning and 
the values of the tuning parameters for various hardware have been manually exported into the production code. 
The implementation requires autotuning to maintain performance portability across GPUs (see Table~\ref{tab:benchmarks-portability}).
%SB breifly say why the resukts have been poor with offline tuning
%JF The sentence was misleading, I have rephrased it.
Because we were unable to install Xmipp on a system with Tesla K20, we have run the benchmark also on GeForce GTX 680 to get more comprehensive results.

%SB it needs to be made clear that Table 8 and 9 refer to offline tuning - right?
%JF yes, fixed
%SB please use ``Geforce GTX'' in the names in addition to numbers - this will make the table more readable %JF done
\begin{table}
\centering
\small
\begin{tabular}{|l|l|l|l|l|}
    \hline
  		& 2080Ti 	&  1070    	& 750 		& 680 \\
    \hline
    2080Ti	& 100\% 	& 99\%		& 31\%		& 49\% \\
    1070 	& 99\%	 	& 100\%		& 31\%		& 50\%\\
    750 	& 43\%	 	& 67\% 		& 100\%		& 94\%\\
    680 	& 60\%	 	& 72\% 		& 71\%		& 100\%\\
    \hline
\end{tabular}
\caption{Performance portability of 3D Fourier reconstruction with $128 \times 128$ samples. The rows represent GPUs used for offline tuning; the columns represent GPUs used for execution% (for example, the worst performance portability is 31\%, which is performance of a code optimized for GTX 1070 running on GTX 750)
. The percentage shows how performance differs compared to the code using the best combination of tuning parameters (for example, the code tuned for GeForce GTX 1070 and executed on GeForce GTX 750 runs at only 31\% of the speed of the code both tuned and executed on GeForce GTX 750).}
\label{tab:fourier-portability-hw}
\end{table}

%SB it needs to be made clear that Table 8 and 9 refer to offline tuning - right?
%JF yes, fixed
\begin{table}
\centering
\small
\begin{tabular}{|l|l|l|l|l|l|}
    \hline
    	 	& 128x128 & 91x91 & 64x64 & 50x50 & 32x32\\
    \hline
    128x128 	& 100\%   & 100\% & 77\%  & 70\%  & 32\%\\
    91x91 	& 100\%   & 100\% & 76\%  & 68\%  & 33\%\\
    64x64 	& 94\%    & 94\%  & 100\% & 91\%  & 67\%\\
    50x50 	& 79\%    & 78\%  & 98\%  & 100\% & 86\%\\
    32x32 	& 65\%    & 67\%  & 80\%  & 92\%  & 100\%\\
    \hline
\end{tabular}
\caption{Performance portability on GeForce GTX1070. The rows represent samples resolution used for offline tuning, the columns represent samples resolution used for execution. The percentage shows relative performance compared to the code autotuned for the used resolution.}
\label{tab:fourier-portability-img}
\end{table}

Detailed performance portability across hardware devices is in Table~\ref{tab:fourier-portability-hw}. The size of the samples inserted into a 3D domain influence the selection of optimal tuning parameters. The tuning has to be repeated for samples of different sizes. Otherwise, suboptimal performance is obtained, as can be seen in Table~\ref{tab:fourier-portability-img}.

%The disadvantage of offline tuning in this scenario is that it is impractical for all combinations of hardware and input sizes. Furthermore, it is also possible to configure interpolation window size, which has an influence on the optimal selection of tuning parameter values. These problems can be solved by introducing online tuning.

%When utilizing online tuning, implementation of the kernel can be changed during the program execution. In order to support it, all CUDA code which directly accesses a GPU has to be removed from a 3D Fourier reconstruction code and replaced with calls to KTT framework. 

\subsubsection{Implementation}

The pseudocode of the reconstruction is shown in Algorithm \ref{alg:fourier}. The output volumes $G$ (Fourier transform of the volume) and $W$ (weights for the 3D voxels) are initialized at the beginning of the computation. In the loop body (lines 4-6), the samples are added into the 3D volume. More precisely, the samples are packed into buffers of a predefined size, and their Fourier transform is computed on a CPU (line 4), copied into GPU memory (line 5) and then tuned GPU kernel is executed to insert the samples into volumes $G, W$ (line 6). The whole algorithm is discussed in detail in~\cite{strelak2019gpu}.

\begin{algorithm}
    \KwIn{$s$}
    \KwOut{$G, W$}
    zero-initialize output volumes $G, W$ in GPU memory\;
    initialize buffer of image's Fourier Transform $s_f$ in GPU memory\;
    \ForEach{$s \in S$} {
        $s_f \leftarrow FFT(s)$ on CPU\;
        upload $s_f$ to GPU\;
        insert projecions of $s_f$ into $G, W$\;
    }
    download $G, W$ to CPU memory\;
    apply weights $W$ to $G$ and perform inverse transform of $G$\;
    \caption{3D reconstruction}
    \label{alg:fourier}
\end{algorithm}

We use the non-blocking dynamic autotuning, which performs both tuning and computation at the same time. %SB this is only true for the non-blocking variant of dynamic tuning as I understood %JF yes, I have written it explicitly
Therefore, all input and output data have to be managed by KTT during the whole program execution. In the loop in line 3, the data are prepared on CPU, then a KTT method \texttt{tuneKernelByStep}, which launches one step of dynamic tuning, is executed. The method selects a new combination of the tuning parameters and executes a tuning manipulator. The tuning manipulator implements lines 5 and 6 of the algorithm. It first copies buffer $s_f$ into GPU memory and then executes a kernel, which inserts samples from $s_f$ into volumes $G, W$. The CPU code is multithreaded, allowing to overlap computation of FFTs with kernel execution. The manipulator uses CUDA streams, so when tuning is finished (and therefore global parallelism is allowed, see Section~\ref{sect:sync}) copying buffers may be executed in parallel with kernel execution and even multiple kernels may be executed in parallel (especially when processing small samples). There is a tuning parameter changing whether atomic writes to output volume in global memory are allowed (see Section~\ref{sect:fourier_tuning_space}). Depending on the tuning parameter value, the tuning manipulator method executes kernel iteratively for each projection (atomic writes are disabled), or just once (processing all samples in buffer $s_f$ in one kernel call). %SB - last sentence unlcear to me %JF is it better now?

\begin{table}
\centering
\small
\begin{tabular}{|l|l|l|l|}
    \hline
     		& best runtime 	& tuning 50 		& tuning full\\
    \hline
    2080Ti 	& 1m40s			& 88\% $\pm$ 3\%	& 54\%\\		% overlay improvement 7%
    1070 	& 5m49s 		& 96\% $\pm$ 2\%	& 79\%\\		% exhaust time 198269241, kernels (offline) 152171363, compilation 46097878, overlay improvement 7%
    750 	& 16m59s 		& 92\% $\pm$ 4\%	& 72\%\\		% exhaust time 704628736, kernels (offline) 659619204, compilation 45009532, overlay improvement 5%
    680		& 15m12s		& 94\% $\pm$ 2\%	& 75\%\\		% exhaust time X, kernels (offline) 522106750, compilation X, overlay improvement 5%
    \hline
\end{tabular}
\caption{The relative performance of dynamically-tuned 3D Fourier reconstruction. The best runtime is measured with \=or\=aculum, \ie{}, the fastest kernel is selected immediately, and no tuning is performed. The relative performance of tuning with searching 50 configurations and with searching the entire tuning space is measured as a percentage of the best runtime. Results for "tuning 50" are shown as an average and standard deviation, whereas other results are shown as an average only (their performance is very stable across multiple executions).}
\label{tab:fourier-dynamic}
\end{table}

\subsubsection{Evaluation}

We have designed an experiment demonstrating the usability of dynamic autotuning with the 3D Fourier reconstruction. We have used a real-world setup, performing reconstruction of the Brome Mosaic Virus \cite{wang2014a} (EMPIAR entry 10010), processing 1,826,160 samples in resolution $156 \times 156$. GPU kernels are processing 1500 samples at once~\cite{strelak2019gpu}; therefore, about 1280 kernels are executed to solve the reconstruction (the actual number can be slightly higher due to a small percentage of void samples). All experiments with different GPUs have been performed on a desktop machine with Intel Core i7-8700. 

In our experiment, the tuning is performed at the beginning of the computation, when both used hardware and sample size are known. The performance of the dynamically tuned code is compared to the performance of code with \=or\=aculum (\ie{}, when the optimal tuning configuration obtained by the offline tuning using exhaustive search is known at the beginning of the computation). We have measured dynamically tuned code in two settings. First, we let KTT perform 50 search steps with random search and then continue with the fastest kernel explored. Second, we perform the exhaustive search and continue with the optimal configuration. As the random search was used, the experiment has been repeated 100 times. Results of this experiment are shown in Table~\ref{tab:fourier-dynamic}. As we can see, the performance penalty of  dynamic tuning is smaller than the performance penalty we get for a code that was tuned offline for a different hardware device (see Table~\ref{tab:fourier-portability-hw}) or different input size (see Table~\ref{tab:fourier-portability-img}). The performance obatined with dynamic tuning  ranges between 88\% and 96\% of the performance of code with \=or\=aculum when 50 configurations are searched, whereas the code mistuned for different GPU can perform at 31\% of \=or\=aculum in the worst case (see Table~\ref{tab:fourier-portability-hw}) and the code tuned for different input size can perform at 32\% of \=or\=aculum in the worst case (see Table~\ref{tab:fourier-portability-img}). %The upper-bound for dynamic tuning is an offline tuned implementation, where offline tuning has been performed at the same hardware and input size. Comparing to implementation tuned offline for different hardware or input size, we can observe how much time the online tuned implementation needs to outperform offline tuning.

We further analyze the overhead of dynamic autotuning. Obviously, the more executions of the kernel (in our case, the more samples used to reconstruct the 3D volume), the less overhead of dynamic autotuning. Therefore, for more complicated reconstructions, the performance of dynamically tuned code is closer to the code using the \=or\=aculum, whereas trivial reconstruction may suffer from dynamic tuning overhead. Adding more work per kernel (in our case using larger samples) decreases relative overhead of the compilation, but not the overhead caused by the execution of slower kernels and synchronization.

In our experiment, the JIT compiler runs for 45.5\,seconds when the full tuning space is searched. It introduces significant overhead in the experiment with GeForce RTX 2080Ti, as the GPU is very fast -- the whole reconstruction with \=or\=aculum is finished in 1\,minute 40\,seconds. With all GPUs, some slowdown is caused by executing slow kernels. The performance of average kernel is at 45\% of the fastest kernel for RTX 2080Ti, 69\% for GeForce GTX 1070, 46\% for GeForce GTX 750 and 52\% for GeForce GTX 680. The good average performance on GeForce GTX 1070 improves the high relative speed of dynamic tuning with 50 explored configurations.

We have also measured the overhead of enforced global synchronization. Recall that in such case, the tuning manipulator copies input samples to the GPU and executes GPU kernels without the overlay with another manipulator instance. The overhead is small for 3D Fourier reconstruction: for kernels executed with enforced global synchronization, it is 7\% for GeForce RTX 2080Ti and GeForce GTX 1070, and 5\% for GeForce GTX 750 and GeForce GTX 680. The global synchronization is enforced when kernels are tuned, but is not needed when tuning is finished.
% --  \ie{}, only 50 out of 1280 kernel executions are slowed by the measured overhead in our setup searching 50 tuning configurations. %SB last sentence unclear %JF better now? %SB not really; I tried to rephrase the last part of the sentence - please check %JF OK
Thus, in our setup, out of the total 1280 kernel executions,  only those 50 launched by the tuning manipulator were slowed down.

%To conclude, we can say that in this example, dynamic tuning is able to overcome performance losses caused by a mistuned code (\ie{}, a code tuned for different hardware or sample resolution) even if the reconstruction program runs in minutes and the number of reconstruction kernel invocations is not high. %SB rewrite ? %JF is the sentence below better?
To conclude, even if the reconstruction program runs in minutes only, dynamic tuning is able to reach better performance than offline tuning in the case offline tuning was performed for different hardware, or different input size.

%\begin{figure}[t]
%    \centering
%    \includegraphics[height=.5\hsize]{figs/fourier-online-2.pdf}
%    \caption{Performance of online tuned 3D Fourier reconstruction.}
%    \label{fig:fourier-perf}
%\end{figure}

%For this experiment, we have used images of resolution $83 \times 83$ pixels and we configured dynamic tuning to perform tuning during first 50 kernel executions. As the search method is stochastic, 100 individual tunings were performed, and the results are shown for the best-performing one, worst-performing one and median. The performance is measured in millions of input image pixels added to the 3D space per second. As can be seen in Figure~\ref{fig:fourier-perf}, the performance of the online tuning is poor during approximatively first 25 seconds, during which the new kernels are compiled and tested. Afterward, the overhead of the initial tuning process is amortized quite quickly. The original code version tuned for Maxwell GPU and images of resolution $128 \times 128$ is outperformed after 42-52 seconds. The code tuned for GeForce GTX 1070 and image resolution $128 \times 128$ is outperformed after 92-243 seconds. The original code tuned offline for GeForce GTX 1070 and image size $83 \times 83$ shows an upper-bound for the test. It cannot be achieved by dynamic auto-tuning, as all combinations of tuning parameters are explored in offline tuning, but its overhead is not included in the measurement.

\subsection{Dynamic tuning of the benchmark set}
The suitability for dynamic tuning for all benchmark can be estimated analytically. We can compare the performance of the best kernel with the average performance of all kernels produced by the tuning space, which allows us to compute the overhead caused by executing slower kernels. %Such analysis allows us to show how difficult is to amortize dynamic tuning (\ie{} how many times the kernel has to be executed to amortize execution of slow kernels explored during autotuning proces). 
%SB unclear: only the realtive compilation overhead cannot be estimated?
%JF yes, I believe absolute overhead has not much sense here, as we don't know the absolute runtime of kernels without knowing the input
We cannot evaluate the relative overhead of kernel compilation, as it depends on application workload (large kernel input prolongs kernel runtime, whereas compilation time remains the same). We also cannot consider the overhead caused by the enforced global synchronization during tuning as it is highly application-dependent if overlapping of manipulators can be leveraged. 
%SB split this sentence  - please check and rephrase? %JF OK
%However, we consider the overhead of slow kernels as the most significant one, as very slow kernels can be created within the tuning space, whereas enforced synchronization limits mainly fast kernels which benefit from concurrent execution and compilation overhead can be hidden by longer kernel runs.
%SB it is still unlcear what the message of the first part of next sentence is? %JF I tried to formulate it differently
The performance penalty of enforced synchronization, as well as kernels compilation, is similar for all tuning variants, whereas the performance penalty of slow kernels can be much higher (some tuning variants can produce orders of magnitude slower kernels). Therefore, we consider the overhead of very slow kernels as the most significant one. 

%SB is the following sentence a correct description of what is done here? If yes, the paragraph below should be rewritten accordingly
In the following we show how to estimate the number $n$ of kernel invocations required in order to amortize the tuning overhead such that the performance of $n$ kernel invocations including the dynamic tuning overhead is a certain fraction of the performance we would have achieved by executing the application using a well-performing kernel $n$ times.
%SB following paragraph needs more improvement .. 
We define a well-performing configuration as a configuration producing a kernel with a performance with which we are satisfied.
%SB in the examples it needs to be clearly said what the actual criteria for ``well-performing'' is; but here it is fine like it is %JF I have added clarification about concept of a well-performing configuration -- it is really a theoretical concept, which allows us to compute, how much kernel execution we need to perform, but this has no direct application in all cases as the well-performing configuration cannot be detected prior to knowledge of the whole tuning space
%(it can be defined, \eg{}, as 95\% of the performance of the best-performing kernel configuration)
 Note that the well-performing configuration can be easily determined with some benchmarks (\eg{}, when defined as a percentage of relevant hardware theoretical peak), but it can be also virtually impossible to identify a well-performing configuration until the whole tuning space is searched (\eg{}, when defined as a configuration reaching a certain fraction of the best configuration performance). In this section, we use a well-performing configuration as a theoretical concept, which is used to determine the number of steps needed to amortize overhead of dynamic tuning. %SB I suggest to skip the following sentence
%SB We do not need it to be computable during an application execution.

%SB we could define T_dyn as the accumulated runtime of s kernel invocations with perf. t_avg + (n-s) invocations with perf. t_well
%SB and T_stat as the accumulated runtime of n kernel invocations with performance t_well
%SB then p is defined as T_dyn/T_stat
Let the application with \=or\=aculum be such an application where a well-performing configuration is known at the beginning of its execution (\eg{}, obtained by previously performed offline tuning). Let the required performance of the dynamically tuned application relative to the performance of the application with \=or\=aculum be 
%since this is required/relative performance it would be better to use ``r'' instead of ``p'' %JF I used "rp", to not collide with relative performance "r" used in eq:samples
$\mathit{rp}$ (so $\mathit{rp}=1.0$ means that the dy\-na\-mi\-cally tuned application runs at the same speed as the application that uses 
%SB? do you ment the best-performing configuration found offline with exhaustive search? %JF it must be a well-performing configuration, as otherwise we cannot compute number of required dynamic tuning steps (we would need to perform exhaustive search)
some well-performing kernel found during offline tuning). Let the number of tuning steps be $s$, the average runtime of kernels within the tuning space be $t_{avg}$ and the runtime of 
%SB probably t_well should be replaced by t_best? %JF see above
the well-performing kernel be $t_{well}$. 
Then, an average\footnote{Here, $\mathit{rp}$ is computed for the average situation with $s$ tuning steps and random search. Obviously, the tuning time may be different from $s \cdot t_{avg}$ in particular executions.} value of $\mathit{rp}$ is computed as:
\begin{equation}
  \mathit{rp} = \frac{s \cdot t_{avg} + (n-s) \cdot t_{well}}{n \cdot t_{well}}
  \label{eq:performance}
\end{equation}

The average number of kernel invocations $n$ required to reach relative performance $\mathit{rp}$ can be estimated as:
\begin{equation}
  n = \frac{\mathit{rp} \cdot s \cdot (\frac{t_{avg}}{t_{well}}-1)}{1-\mathit{rp}}
  \label{eq:invocations}
\end{equation}

%SB unclear what is meant - rewrite? %JF better? %SB yes!
For example, if the average kernel has runtime $t_{avg} = 10\,ms$, the well-performing kernel has runtime $t_{well}=5\,ms$, we perform $s=100$ tuning steps and we want to reach relative performance $\mathit{rp}=0.9$ (\ie{}, dynamic autotuning reach 90\% of the performance of an application with \=or\=aculum), we need to perform 900 kernel invocations (including those used for tuning). 

The real amortization of dynamic autotuning depends on the number of tuning steps required to find a well-performing kernel. When random search is used, the number of required tuning steps can be computed as follows. Let $r$ be the ratio of well-performing configurations in the tuning space and $p$ be the required probability of finding a well-performing configuration. The number of tuning steps $s$, which leads to reaching the well-performing configuration with probability $p$, can be computed as

\begin{equation}
  s = \log_{1-r}(1-p)
  \label{eq:samples}
\end{equation}

For example, if the ratio of well-performing configurations is $r=0.01$, then we need to explore 230 configurations in order to reach a well-performing configuration with probability $p=0.9$.

\begin{table*}
\centering
\small
\begin{tabular}{|l|r|r|r|r|r|r|r|}
    \hline
    Benchmark 	& 2080Ti 	& 1070 		& 750 		& K20 		& Vega56	& E5-2650 	& 5110P \\
    \hline
    BiCG	& 10,383 	& 9,425		& 33.090	& 43,552	& 42,499	& 32,338	& 516,783\\
    2D Convolution & 265,297	& 98,966	& 197,550	& 165,783	& 99,087	& 7,211		& 3,435\\
    Coulomb 3D  & 17,305	& 16,346	& 4,911		& 5,289		& 117*		& 150		& 631\\
    GEMM	& 20,309	& 151,564	& 764,485	& 205,122	& 18,782*	& 384,309	& 3,106,384\\
    GEMM batched & 2		& 2		& 110		& 214		& 440		& 2,341		& 1,214\\
    Hotspot	& 4,314		& 4,467		& 3,309		& 5,635		& 1,489*	& 3,926		& 7,346\\
    Transpose   & 9,398		& 347		& 2,998		& 1,347		& 140,177	& 5,129		& 60,688\\
    N-body	& 7,539		& 33,553	& 2,531		& 20,694	& 554*		& 2,472		& 1,669,559\\
    Reduction	& 646		& 78		& 40		& 218		& 2,198		& 1,650		& 19,425\\
    3D Fourier	& 2,239 	& 830		& 3,123		& N/A		& N/A		& N/A		& N/A \\
%    Sort 	&		&		&		&		&		&		& \\
    \hline
\end{tabular}
\caption{The number of kernel invocations required to hide overhead of slow kernels execution. The goal is to find a kernel within 95\% of the optimum with 90\% probability and decrease tuning overhead under 10\% of the runtime. Benchmarks on Radeon RX Vega56 marked with * are running with smaller tuning space due to ROCm instability.}
\label{tab:benchmarks-dynamic}
\end{table*}

Using Equations~\ref{eq:invocations} and~\ref{eq:samples}, we can compute the number of kernel invocations needed to hide overhead caused by executing slow kernels during dynamic tuning. We have set up the following experiment. We define a well-performing configuration as a configuration, which leads to at least 95\% of the best configuration performance. Using data gathered from the offline tuning of our benchmark set with exhaustive search, we have computed the number of tuning steps required to find a well-performing configuration with probability 0.9 (using Equation~\ref{eq:samples}). Then, we have computed the number of kernel executions required to decrease dynamic tuning overhead under 10\%  (\ie{} to obtain at least 90\% of the performance we would have with \=or\=aculum). The results are given in Table~\ref{tab:benchmarks-dynamic}.

Table~\ref{tab:benchmarks-dynamic} demonstrates that dynamic tuning is feasible even for short program executions (with thousands or tens of thousands of kernel calls) with multiple benchmarks, such as BiCG, Direct Coulomb Summation, Batched GEMM, Hotspot, Transpose, N-body, Reduction and 3D Fourier Reconstruction. Longer execution is needed for 2D Convolution and GEMM benchmarks. This test also shows some interesting differences between the hardware devices used in the test. For example, autotuning of OpenCL code for a CPU is similarly demanding as for GPUs with many benchmarks, whereas it is much harder on the MIC (Xeon Phi) in multiple cases. There are also some benchmarks where different hardware performs highly differently. For example, with Batched GEMM on GeForce GTX 1070, 304 configurations out of 424 are within 95\% of the optimum and the average kernel performance is at 95\% of the optimum, so it is very easy to find a well-performing kernel and to amortize tuning overhead. With GeForce 750, only 40 configurations produce well-performing kernels, and the average kernel performance is within 62\% of the best one, so tuning is harder  than for GeForce GTX 1070. With Xeon E5-2650,  only three tuning configurations result in a well-performing Batched GEMM kernel, and the average kernel performance is at 51\% of the best kernel, so searching a well-performing kernel and amortizing the tuning overhead is significantly harder on the CPU. Interesting differences between GPU and CPU/MIC can be seen in 3D Coulomb Summation benchmark, where tuning for GPUs is harder. Not only is the number of well-performing kernels different (\eg{}, 110 for Xeon E5-2650 and 28 for GeForce GTX 1070), but a more significant difference is the average performance -- it is much lower for all GPUs (\eg{}, 5\% of the best one on GeForce GTX 1070 vs. 57\% on Xeon E5-2650). The poor average speed on GPU is caused by huge register spilling when high unrolling of the inner loop is used. Although it would be easy to remove these slow-performing configurations from the tuning space, we decided to keep the space as it was designed when the benchmark was developed instead of adding a posteriori information for tuning.

\section{Conclusion and Future Work}
\label{sect:conclusion}

In this paper, we have introduced the Kernel Tuning Toolkit  -- an advanced autoning framework for OpenCL and CUDA applications. Using KTT allows expert programmers to configure applications for offline tuning and dynamic tuning based on arbitrary user-defined code optimization parameters. We have also developed a set of ten benchmarks covering important HPC application areas and demonstrated that autotuning with KTT allows to produce highly efficient implementations (often close to the theoretical peak of the hardware) of these benchmarks for different hardware architectures including CPUs, Xeon Phi co-processors and GPUs. In our experimental evaluations we also demonstrate that autotuning for different architectures is key for performance portability. Moreover, we have shown that rationally-de\-sig\-ned tuning spaces are often small enough to be searched during application runtime, making dynamic tuning feasible for a subset of the considered benchmarks. We have demonstrated with two different applications that dynamic tuning outperforms offline tuned implementations quickly if some performance-relevant aspects, such as a the size of data structures, change. Moreover, we have shown that our framework can be integrated into production software, supporting multi-threading, overlapping execution of host and device code with memory copies, and utilizing simultaneous kernel execution.

In future work, we would like to focus on the further development and integration of advanced search strategies. We believe that it is possible to accelerate dynamic tuning by gathering properties of the tuning space from previous tuning runs, \eg{}, determine the relative impact of tuning parameters on performance by analyzing of profiling data. Using more efficient search methods would make dynamic autotuning feasible also for larger tuning spaces with a small number of well-performing configurations. Another line of research will focus on advanced dynamic strategies that can detect when the performance of an application degrades and that can then automatically trigger dynamic re-tuning of the code.
%When gathering profiling data during application runtime, we believe it is also possible to compute the efficiency of the tuned kernels. Then, when some performance-relevant characteristic changes (\eg{}, size of the input), KTT may trigger dynamic tuning automatically to seek  for better-performing kernel.

Another planned improvement targets the generation of tuning spaces. Currently, KTT first generates the whole tuning space and then prunes it based on the constraints given. We plan to speed-up tuning space generation similarly as it has been done in the ATF framework~\cite{rasch2018atf}.

Furthermore, we plan to investigate the possibilities of connecting KTT with a compiler-based approach. By introducing a DSL for optimizations, the programmer would need to only implement advanced optimizations (such as changing memory layout or the algorithm), whereas simpler optimizations (such as vectorization or loop blocking) would be generated automatically by the compiler.

The vast amount of autotuning results, especially when coupled with profiling counters, can be used by the community to compare behavior and efficiency of different HW architectures, study effects of different code optimizations, or to develop new search strategies. Therefore, we plan to set up a public database containing tuning results with profiling counters and update this database with any new hardware or benchmark available.

Last but not least, KTT has been designed to be independent of the concrete API used for accelerated kernels (\eg{}, OpenCL or CUDA). It is, therefore, possible to add broader support for APIs, for example, Vulcan support would extend potential applications of KTT towards computer graphics. With non-blocking dynamic tuning, it would be possible to alter shaders at runtime without significant drop of frame rate.

\section*{Acknowledgements}
\small
The work was supported from European Regional Development Fund-Project "CERIT Scientific Cloud" (No. CZ.02.1.01\-/0.0/0.0/16\_013/0001802).
The project that gave rise to these results received the support of a fellowship from ”la Caixa” Foundation (ID 100010434). The fellowship code is LCF/BQ\-/DI18\-/11660021.
This project has received funding from the European Union’s Horizon 2020 research and innovation programme under the Marie Skłodowska-Curie grant agreement No. 713673.
The Spanish Ministry of Economy and Competitiveness through Grants BIO2016-76400-R(AEI/FEDER, UE).
“Comunidad Autónoma de Madrid” through Grant: S2017\-/BMD-3817.
%European Union (EU) and Horizon 2020 through grant INSTRUCT - ULTRA (INFRADEV-03-2016-2017, Proposal: 731005).
%European Union (EU) and Horizon \linebreak2020 through grant EOSC Life (INFRAEOSC-04-2018, Proposal: 824087).

\bibliographystyle{plain}
\bibliography{fila}

\end{document}